\documentclass[]{jfm}
\paperheight\pdfpageheight 
\paperwidth\pdfpagewidth   

\newcommand{\vect}[1]{\boldsymbol{#1}} 
\newcommand{\nablaperp}{\nabla_{\!\!\perp}}    
\newcommand{\I}{\mathrm{i}}            
\newcommand{\e}{\mathrm{e}}            

\makeatletter
\newcommand*{\hyperlinkcite}[1]{\hyper@link{cite}{cite.#1}}
\makeatother

\newcommand{\hzero}{H_{0}}
\newcommand{\hinfty}{H_\infty}

\newcommand{\upertvec}{\vect{u}^\prime}
\newcommand{\upertperp}{\vect{u}_\perp^\prime}
\newcommand{\upert}{u^\prime}
\newcommand{\vpert}{v^\prime}
\newcommand{\wpert}{w^\prime}

\newcommand{\dpdz}{\left.\frac{\partial p}{\partial z}\right|_{H_{0}}}

\newcommand{\Fr}{\mathrm{Fr}}
\newcommand{\Zbase}{f_0}
\newcommand{\Ybase}{q_0}

\newcommand{\BS}{b_{0}}

\usepackage[normalem]{ulem}

\lefttitle{E. Zuccoli, E.J. Brambley and D. Barkley}
\righttitle{Journal of Fluid Mechanics}

\title{A free-surface-only closure model for linear waves on general deep-water flows}

\author{Emanuele Zuccoli\aff{1}, Edward J.~Brambley\aff{2}\aff{,3} \and Dwight Barkley\aff{2}}

\affiliation{\aff{1} School of Mathematics, Cardiff University, Cardiff, CF24 4AG, United Kingdom
\aff{2} Mathematics Institute, University of Warwick, Coventry, CV4 7AL, United Kingdom
\aff{3} WMG, University of Warwick, Coventry, CV4 7AL, United Kingdom}

\corresau{Edward J.~Brambley, \email{E.J.Brambley@warwick.ac.uk}}

\begin{document}
\maketitle

\begin{abstract}

We present a novel, spatially two-dimensional (2D) set of equations to study the propagation of linear deep-water surface waves over a general steady three-dimensional (3D) background free surface flow.
No assumptions of a flat background free surface, nor of an irrotational background flow, are made.
The resulting model is not only a significant theoretical simplification, but also results in orders of magnitude faster computations.
The linearized Euler equations are evaluated on the free surface of the base flow, and a closure condition is proposed to account for the vertical derivatives at the free surface.
This generalizes a recent result for purely rotating background flows \hyperlinkcite{Zuccoli2025}{(Zuccoli, Brambley \& Barkley, 2025, Phys.\ Rev.\ Fluids~10, 024801)}.
The final model consists of five coupled first order partial differential equations (PDEs) to be solved on the 2D free surface, involving five unknowns: the horizontal and vertical perturbation velocities; the free surface perturbation height; and unexpectedly the gradient of perturbation pressure with depth at the surface.
Two test problems are used to validate the model: a uni-directional vortical depth-varying base flow with a flat free surface; and perturbations to a travelling Gerstner wave solution.
Eigenvalues and eigenfunctions of the linearized Euler equations are computed and compared with those of the model. Results show remarkable agreement between the two.
Our study finds for the first time, to the best of our knowledge, that sufficiently steep two-dimensional Gerstner waves are unstable.
\end{abstract}

\begin{keywords}
deep-water surface waves, free surface flows, reduced models, Gerstner waves
\end{keywords}

\section{Introduction}

The propagation of free-surface waves over non-trivial background flows has been extensively studied over the past several decades, motivated both by their rich dynamics and their relevance to a wide range of physical applications. The background flow, or base flow, can profoundly affect the propagation and stability of free-surface waves, leading to phenomena that are absent for a quiescent background flow. Examples include the stability of free surfaces over parallel, unidirectional shear flows \citep{caponi91, Longuet-Higgins98}, the stability of free-surface swirling flows \citep{acheson,tophoj+others,jansson+others,bach+others,mougel-2014,mougel-2015,mougel-2017, zuccoli+brambley+barkley-2022}, and the modelling of waves generated by wind-driven currents and wakes \citep{Dimas94, Young2013}. 

In this work, we consider linear surface waves in the form of small perturbations to a prescribed base flow. We neglect both viscosity and surface tension. Even in this simple case, the resulting linear free-surface problem is, in general, three-dimensional (3D) in space and therefore frequently computationally demanding. The difficulty is particularly pronounced when the base flow has a deformed free surface and non-zero vorticity, and when the horizontal domain is effectively unbounded. In the latter case, numerical calculations require appropriate treatment of the computational boundaries to prevent artificial reflection of waves. The simultaneous presence of these features limits the possibility of separating the underlying spatial eigenvalue problem, as is possible in many simpler free-surface flows, and consequently greatly increases the computational cost of obtaining normal-mode solutions.

For potential flows, the three-dimensional free-surface problem can be reduced to a formulation involving quantities defined only on the free surface. This reduction is based on the Dirichlet-to-Neumann (DtN) operator, which maps the value of the velocity potential on the free surface to its normal derivative there, thereby taking into account the fluid interior without explicitly resolving the flow throughout the depth. Following the pioneering work of \citet{zakharov1968stability}, surface formulations based on the DtN operator have been developed extensively
\citep{craig1993numerical, milewski+keller}. 
For linear waves over a quiescent fluid of constant depth $\hinfty$, the resulting surface equation can be written as
\begin{equation}
\frac{\partial^2\phi}{\partial t^2}+gG_0\phi=0,
\label{eq:flat_potential}
\end{equation}
where $\phi$ is the velocity potential of the perturbation at the free surface and $G_0$ is the DtN operator for a flat surface,
\begin{equation}
    G_0=(-\nabla_\perp^2)^{1/2}\tanh\left( \hinfty(-\nabla_\perp^2)^{1/2}\right),
\label{eq:G0}
\end{equation}
with $\nabla_\perp=(\partial_x,\partial_y)$, where $(x,y)$ are horizontal coordinates.
In one horizontal spatial dimension, the operator $G_0$ is commonly written $G_0= - \I\partial_x\tanh\left(-\I\hinfty\partial_x \right)$.
In Fourier space, $G_0$ acts as multiplication by $|\vect{k}|\tanh(|\vect{k}|\hinfty)$, where $\vect{k}$ is the wave vector, thus giving the familiar dispersion relation $\omega^2=g|\vect{k}|\tanh(|\vect{k}|\hinfty)$. 
In the infinite-depth limit, $\hinfty\to\infty$, the DtN operator reduces to 
$G_0=(-\nabla_\perp^2)^{1/2}$ 
and the dispersion relation becomes $\omega^2=g|\vect{k}|$.

\citet{torres+coutant+dolan+weinfurtner} generalized this approach to study surface waves on a background potential vortex $\bf U_0$. They assumed that the background free surface remained flat, thus obtaining
\begin{equation}
\begin{aligned} &
\hat{D}^{2}_{t} \phi + gG_0\phi = 0,
\end{aligned}
\label{eq:Torres}
\end{equation}
where $\phi$ is again the velocity potential for linear perturbations, now with respect to a non-zero background flow, $\hat{D}_t = \partial_t + \vect{U_0\! \cdot\! \nablaperp}$ is the convective derivative with respect to the background flow $\vect{U_0}$, and $G_0$ is again the general DtN operator defined by \eqref{eq:G0}.

The situation is more difficult for waves propagating on vortical background flows. In this case, the classical DtN approach is not applicable and does not provide a closure to obtain a surface-only model. Interior vorticity provides additional degrees of freedom, so that the evolution of perturbation fields cannot in general be determined from their values on the free surface alone, even in the linear case. An additional assumption concerning the behaviour of linear surface wave in the vicinity of the free surface is therefore required.

Recently, \citet{Zuccoli2025} proposed such a surface-only approach to study the stability of deep-water surface waves on vortical axisymmetric background flows. The closure was motivated heuristically by the behaviour of linear surface-wave perturbations near the free surface, rather than being derived as a rigorous consequence of the governing equations.
Their model accounted for both free-surface deformation and the unbounded lateral domain, thus addressing the drawbacks outlined above. Moreover, the resulting model consists of four coupled first-order PDEs and is therefore readily amenable to standard numerical treatment.

In the present paper, we generalize the reduced surface-only model obtained by \citet{Zuccoli2025} to non-axisymmetric, depth-dependent, three-dimensional (3D) base flows. The resulting surface-only equations are derived in \S\ref{sec:maths_model}. To assess the validity and computational efficiency of the surface-only model, we consider two test problems of increasing complexity in \S\ref{sec:validation_model1} and \S\ref{sec:validation_model2}. For both test problems, we compare modal solutions of the surface-only equations with modal solutions of the full-depth linearized Euler equations. The first test problem considers a unidirectional, vertically varying base flow with a periodic structure in the spanwise direction. The corresponding free surface remains flat, but the base flow has non-zero vorticity and gives rise to a non-trivial eigenvalue problem for modal solutions. The second test problem considers the stability of Gerstner waves. In this case, the base flow varies in both the vertical and streamwise directions, has non-zero vorticity, and produces a non-zero free-surface deformation. For both test problems, the surface-only model reproduces the eigenvalues and eigenfunctions of the full-depth problem surprisingly well, and, in the case of Gerstner waves, it captures the onset of instability. Finally, conclusions and opportunities for future research are presented in \S\ref{sec:conclusions}.

\section{Mathematical Formulation}\label{sec:maths_model}

\subsection{Governing equations and base flow}

We assume that viscosity is negligible and hence that the fluid flow is governed by the incompressible Euler equations,
\begin{align}
&\frac{\partial\vect{U}}{\partial t} + \vect{U\cdot \nabla}\vect{U} = - \bnabla P  - g\vect{\hat{z}}, &&
\vect{\nabla \cdot U} = 0,
\label{equ:euler}
\end{align}
where $\vect{U} = (U,V,W)$ is the velocity, $P$ is the pressure scaled by the constant density $\rho$, $\vect{\hat{z}}$ is a unit vector in the vertical direction, and $g$ is the acceleration due to gravity.
The fluid is contained between an upper free surface at $z = H(x,y)$, in general varying with the horizontal spatial coordinates $\vect{x}_{\perp}=(x,y)$, and a bottom boundary at $z=-\hinfty$, where $\hinfty$ is constant. We are in principle interested in the infinite-depth case, $\hinfty= \infty$, but for numerical simulations $\hinfty$ is finite, in which case the fluid satisfies no penetration through the bottom boundary: $\vect{U\cdot\hat{z}} = 0$ at $z=-\hinfty$.
The flow must satisfy kinematic and dynamic boundary conditions at the free surface. We neglect surface tension and assume the fluid above the free surface to be dynamically passive, and in particular, to have a constant pressure $\bar{P}$.  Together, these give the boundary conditions
\begin{subequations}\label{equ:full_bcs}\begin{align}
P = \bar{P} \quad \text{and} \quad
\vect{U\cdot\hat{z}} = \frac{\partial H}{\partial t} + \vect{U_{\perp}\cdot \nabla_{\perp}} H
\quad \text{at} \quad
z = H,
\tag{\theequation a,b}
\end{align}\end{subequations}
where $\vect{U}_{\perp} = (U,V)$ and $\nabla_{\perp} = (\partial_x, \partial_y)$ are the horizontal velocity and horizontal gradient, respectively.
We consider base flows $\vect{U}_{0} = (U_0, V_0, W_0)$, $P_0$ and $H_0$ satisfying the time-independent equations form of \eqref{equ:euler} and \eqref{equ:full_bcs}.

\subsection{Linearized equations}

We are interested in the behavior of infinitesimal perturbations to the base flow. 
Let $\upertvec = (\upert,\vpert,\wpert)$ be the perturbation velocity, $p$ be the perturbation pressure, and $h$ the perturbation height, so that the total velocity is $\vect{U} = \vect{U}_0 + \upertvec$, the total pressure is $P = P_0 + p$, and the total fluid elevation is $H = H_0 + h$. (We reserve $\vect{u}$ for the perturbation velocity restricted to the free surface that will appear shortly.) Assuming the perturbations to be small, the perturbation dynamics are given by linearizing the incompressible Euler equations~\eqref{equ:euler} and boundary conditions~\eqref{equ:full_bcs} about the base flow, resulting in the following
\begin{subequations}\begin{align} &
D_{t}\upertvec + (\upertvec\cdot\nabla)\, \vect{U}_{0} + \nabla p = 0, \label{equ:horiz_mom_eq} \\ &
\nabla\cdot\upertvec = 0, \label{equ:cont_eq} \\ &
p = g_0 h, \quad \mathrm{on} \quad z = H_0\label{equ:dyn_BC}, \\ &
\wpert = \frac{\partial h}{\partial t} + \boldsymbol{U_0}_{\perp}\cdot\nabla_{\perp}h + \upertperp\cdot\nabla_{\perp}H_0 - \varphi_{0} h, \quad \mathrm{on} \quad z = H_0,\label{equ:kin_BC}, \\ &
\wpert = 0, \quad \mathrm{on} \quad z = -\hinfty, \label{equ:bottomBC}%
\end{align}
\label{equ:linearised_euler}
\end{subequations}
where $D_t = \partial_t + \vect{U}_{0}\cdot\nabla$ is the convective derivative due to the base flow. We have defined two quantities on the free surface:
\begin{align}
g_0 & = g + (\boldsymbol{U_0}\cdot\nabla) \, W_0  \Big|_{H_0}, \\
\varphi_0 & = \left.\frac{\partial W_0}{\partial z}\right|_{H_{0}} - \left.\frac{\partial \boldsymbol{U_0}_{\perp}}{\partial z}\right|_{H_{0}}\cdot\nabla_{\perp} H_0.
\end{align}
Here, $g_0$ is the effective gravity accounting for vertical acceleration of the base flow on the free surface, while $\varphi_0$ results from expressing the base normal velocity at the perturbed free surface $H_0+h$ as an expansion about the base free surface $H_0$. Here and throughout, $|_{H_0}$ means evaluated at the base free surface $z = H_0$. Our interest here is in surface waves in deep water, for which perturbations decay away from the free surface. For infinite depth, the bottom boundary condition becomes a requirement that the perturbations decay as $z\to-\infty$.

\subsection{Surface-only model}
\label{sec:closure}

We seek equations expressed only in terms of quantities defined the base free surface, and hence quantities that depend only on the horizontal coordinates $\vect{x}_{\perp}$, and not depth $z$. First, we let $\vect{u}$ denote the perturbation velocity on free surface
\begin{equation}
    \begin{aligned} &
        \vect{u}(\vect{x}_{\perp}, t) = 
        \upertvec(\vect{x}_{\perp}, H_0(\vect{x}_{\perp}), t).
    \end{aligned}
\end{equation}
The components of $\vect{u}$ are denoted $(u,v,w)$ and we use $\vect{u}_\perp$ to represent the horizontal components $(u,v)$.

We use two identities for derivatives evaluate on the free surface. Let $f(\vect{x}_{\perp}, z, t)$ represent any of the unknowns $u, v, w, p$. Then, 
\begin{subequations}\begin{align}
&
\left(\nabla_{\perp} f \right)|_{H_0} = \nabla_{\perp}(f|_{H_0}) - \nabla_{\perp}H_0\left.\frac{\partial f}{\partial z}\right|_{H_0}, \label{equ:Gradient_surface} \\ &
\left. (D_{t}f) \right|_{H_{0}} = \frac{\partial f|_{H_0}}{\partial t} + \vect{U}_{0_{\perp}}|_{H_0}\!\cdot\!\nabla_{\perp}(f|_{H_0}) + (W_{0}|_{H_0} - \vect{U}_{0_{\perp}}|_{H_0}\!\cdot\!\nabla_{\perp}H_0)\!\left.\frac{\partial f}{\partial z}\right|_{H_0} \!\!\!= \hat{D}_{t}\left(f|_{H_{0}}\right) \label{equ:Dtsurface},
\end{align}\label{equ:relations_unknwons_on_FS}
\end{subequations}
where we have introduced $\hat{D}_t = \partial_t + \vect{U}_{0_{\perp}}|_{H_0}\cdot\nabla_{\perp}$ as the convective derivative operator acting on the free surface. The term $W_{0}|_{H_0} - \vect{U_{0_{\perp}}}|_{H_0}\cdot\nabla_{\perp}H_0$ in
\eqref{equ:Dtsurface} is zero due to the kinematic boundary condition \eqref{equ:full_bcs} for the base flow.

Using identities \eqref{equ:relations_unknwons_on_FS}, the linearized momentum equations \eqref{equ:horiz_mom_eq}, the incompressibility constraint \eqref{equ:cont_eq}, and the
kinematic boundary condition \eqref{equ:kin_BC} evaluated on the free surface $z = H_0$ become
\begin{subequations}\begin{align} &
\hat{D}_t \vect{u}_{\perp}
+ \left(\vect{u}\cdot\nabla\right)\, \vect{U}_{0_{\perp}}|_{H_0}
+ \nabla_{\perp}(p|_{H_0}) - 
\nabla_{\perp}H_0
\dpdz=0,
\label{eq:u_perp} \\ &
\hat{D}_t w 
+ \left(\vect{u}\cdot\nabla\right)\, W_0|_{H_0} + \dpdz = 0,
\label{eq:w} \\ &
\nabla_{\perp}\cdot\vect{u}_{\perp}
+ \left.{\frac{\partial \wpert}{\partial z}}\right|_{H_{0}}
- \nabla_{\perp}H_0\cdot\left.{\frac{\partial \upertvec_{\perp}}{\partial z}}\right|_{H_{0}} = 0, \label{equ:continuity_on_FS}\\ &
w = \hat{D}_t h
+ \left(\vect{u}_{\perp}\cdot\nabla_{\perp}\right) H_0
- \varphi_{0}h,\label{equ:FSBC_perturbations}
\end{align}%
\label{equ:linearised_euler_on_FS_2}%
\end{subequations}%
where we now explicitly separate the equations for the horizontal and vertical momenta on the free surface, equations \eqref{eq:u_perp} and \eqref{eq:w}.

Equations \eqref{equ:linearised_euler_on_FS_2} are not closed because they contain vertical derivatives at the free surface. Following from our previous work \citep{Zuccoli2025}, we propose, but do not formally derive, a closure condition to expresses the vertical velocity gradients that appear in the incompressibility constraint in terms of the time derivative of the vertical pressure gradient.

Equation \eqref{equ:FSBC_perturbations} is the kinematic boundary conditions that the perturbations fields $\upertvec_{\perp}, w', h$ must satisfy on the free surface. Using the dynamic boundary condition \eqref{equ:dyn_BC} to replace $h$ by $p/g_0$ on the free surface, and the boundary condition can be written
\begin{equation}
    \Psi[\upertvec_{\perp}, w', p](\vect{x}_{\perp}, \hzero(\vect{x}_{\perp}), t) = 0,
\end{equation}
where $\Psi$ is the boundary operator
\begin{align} &
    \Psi[\upertvec_{\perp}, w', p] =  w' 
- \left(\vect{u}'_{\perp}\cdot\nabla_{\perp}\right) H_0
- \hat{D}_t \left( \frac{p}{g_0} \right)
+ \frac{\varphi_{0}}{g_0} p,
\label{functional_Psi}
\end{align}
The ansatz we make to close the system is that, for surface gravity waves, the perturbation fields satisfy
\begin{equation}
    \Psi[\upertvec_{\perp}, w', p](\vect{x}_{\perp}, \hzero(\vect{x}_{\perp}) - \epsilon, t) = o(\epsilon).
    \label{eq:ansatz}
\end{equation}
In other words, at distance $\epsilon$ below the free surface, perturbation fields obey the same boundary relationship as on the free surface, to within an error that is smaller than $\epsilon$ as $\epsilon \to 0$. We discuss this assumption below, but first proceed with the manipulations, noting that everything follows from the ansatz. 

From this ansatz,
\begin{equation}
    \left. \partial_z \Psi[\upertvec_{\perp}, w', p](\vect{x}_{\perp}, z, t)\right|_{z=\hzero} = 0
\end{equation}
from which we obtain a relationship between vertical gradients of velocity fields and pressure on the free surface
\begin{equation}
\left.{\frac{\partial \wpert}{\partial z}}\right|_{H_{0}} - \left.{\frac{\partial \upertvec_{\perp}}{\partial z}}\right|_{H_{0}}\cdot\nabla_{\perp} H_0 = 
\hat{D}_{t}\Big(\frac{1}{g_0}\dpdz \Big) - \frac{\varphi_0}{g_0} \dpdz.
\label{equ:closure_final}
\end{equation}
This is our closure condition.
On the left-hand side are the vertical derivatives appearing in the continuity equation on the free surface, \eqref{equ:continuity_on_FS}. 
On the right-hand side are vertical derivatives of pressure on the free surface, $\partial_zp|_{H_0}$, and more specifically, there is a time derivative of $\partial_zp|_{H_0}$. 
The final system of equations is then obtained by substituting the closure condition \eqref{equ:closure_final} into the continuity equation \eqref{equ:continuity_on_FS},
and again using the dynamic boundary condition, this time to replace $p|_{H_{0}}$ by $g_0 h$.
This results in a closed system of equations for the surface variables $\vect{u}_{\perp}, w, h, \partial_zp|_{H_0}$ as functions of $(\vect{x}_{\perp},t)$:
\begin{subequations}\begin{align} &
\hat{D}_t \vect{u}_{\perp} 
+ (\vect{u}_{\perp}, w)\cdot\nabla \vect{U}_{0_{\perp}}|_{H_0}
+ \nabla_{\perp}(g_0 h) - \nabla_{\perp}H_0\dpdz = 0,
\label{equ:model_horiz_momentum_eqs} \\ &
\hat{D}_t w 
+ (\vect{u}_{\perp}, w)\cdot\nabla W_0|_{H_0}
+ \dpdz = 0,
\label{equ:model_vert_momentum_eq} \\ &
\hat{D}_{t}\Big(\frac{1}{g_0}\dpdz\Big) - \frac{\varphi_0}{g_0}\dpdz + \nabla_{\perp}\cdot\boldsymbol{u}_{\perp} = 0, 
\label{equ:model_cont_eq} \\ &
\hat{D}_t h
+ \boldsymbol{u}_{\perp}\cdot\nabla_{\perp}H_0
- \varphi_{0}h - w = 0. \label{equ:model_kin_BC}
\end{align}%
\label{equ:final_reduced_system}%
\end{subequations}
We will use {\em surfaces-only} when referring to equations \eqref{equ:final_reduced_system} and their solutions. In contrast, we will use {\em full-depth} when referring to the full linearized Euler equations~\eqref{equ:linearised_euler} and their solutions.

\subsection{Discussion of the surface closure}

We begin by recapping the closure condition and the resulting surface-only model equations. Expressing the linearized Euler equations on the base free surface in terms of free-surface quantities gives equations \eqref{equ:linearised_euler_on_FS_2}. Vertical derivatives necessarily appear in these equations. The vertical pressure gradient $\partial_z p|_{H_0}$ appears in the horizontal momentum equation as a result of the non-flat free surface, and in the vertical momentum equation through the usual force balance. Similarly, the surface continuity equation contains vertical derivatives, both as a consequence of its standard form and as a result of the non-flat free surface. While none of these vertical derivatives is known from the surface values of pressure or velocity, the closure condition  \eqref{equ:closure_final} relates the derivatives. In particular, the closure condition relates the unknown vertical derivatives in the continuity equation to a time derivative of the free-surface pressure gradient $\partial_z p|_{H_0}$. This leads to two simultaneous features of the surface-only model \eqref{equ:final_reduced_system}: the gradient $\partial_z p|_{H_0}$ becomes an independent dynamical variable, and the continuity equation \eqref{equ:model_cont_eq} becomes a time-evolution equation governing its dynamics. 
The resulting surface-only model equations are first-order in time with dependent variables $\vect{u}_{\perp}, w, h,$ and also $\partial_zp|_{H_0}$.

The closure condition is central to the surface-only model and it merits further discussion. We recall the simplest possible case of linear perturbations about a motionless, infinitely deep base flow. The kinematic and dynamic boundary conditions combine to give $w' = \partial_t p/g$ on the free surface. Using a velocity potential, it is straightforward to show that this relationship holds not only on the free surface, but throughout the fluid. Thus, one may differentiate to obtain $\partial_z w' = \partial_t (\partial_z p/g)$, which is the closure condition \eqref{equ:closure_final} evaluated in the simplest case. 
More generally, as long as the base flow is potential flow and the base free surface is flat, it is possible to show that the closure condition
$\partial_z w' = \hat{D}_t \left( \partial_z p/g \right)$ holds exactly.
The point is that this relationship between the vertical velocity gradient and the time derivative of the vertical pressure gradient is a basic feature of linear gravity waves on a flat free surface.

For the general case of surface gravity waves propagating over vortical base flows, we have been unsuccessful in deriving the closure condition under suitable assumptions other than the ansatz \eqref{eq:ansatz}. We therefore do not claim that the closure condition is exactly valid in general, or even that it holds in some asymptotic sense, other than vanishing vorticity and a flat free surface. The difficulty in deriving such a closure reflects a fundamental difference between potential and vortical flows. In potential flow, the structure of the equations allows the behaviour of the velocity field in the fluid to be determined from data on the free surface through the Dirichlet-to-Neumann map. For a general vortical flow, however, the interior vorticity represents additional bulk degrees of freedom that cannot in general be recovered from free-surface data alone. Consequently, there is no analogous closure based solely on the values of the fields at the free surface without additional assumptions about the interior flow. Ansatz \eqref{eq:ansatz} and the resulting closure condition  \eqref{equ:closure_final} should therefore be understood as a hypotheses specifically about linear surface gravity-wave solutions in the immediate vicinity of the free surface, rather than as a general property of vortical flows or of solutions of the linearized Euler equations.

\cite{zuccoli-2023} considered the closure condition in the cylindrical axisymmetric case with no axial flow and showed computationally that the closure condition is well satisfied in deep water for waves on axisymmetric vortices. In the examples considered, the base flows and perturbation fields were both vortical, and the base free surfaces were highly deformed by centrifugal acceleration. In the present work, we have extended the closure condition and surface-only model to more general cases. We thus seek to investigate a broader class of examples and to determine under what conditions the closure is approximately satisfied and the surface-only equations produce quantitatively useful solutions.

\section{Validation 1 --- waves on a flat shear flow}
\label{sec:validation_model1}

We validate the surface-only model \eqref{equ:final_reduced_system} by comparing eigenvalues and eigenmodes from the model with those obtain from the full-depth equations \eqref{equ:linearised_euler}. 
In this section, we consider a relatively simple test case in which the base velocity field is unidirectional and the base free surface is flat. Specifically we consider $\vect{U}_{0}(\vect{x}_{\perp}, z) = U_{0}(y, z)\vect{\hat{x}}$, where $\vect{\hat{x}}$ is the unit vector in the $x$-direction. Irrespective of the form of $U_{0}(y, z)$, the base free surface is flat, so we set $H_0(\vect{x}_{\perp}) = 0$, and the base pressure field is hydrostatic and given by $P_0(\vect{x}_{\perp}, z) = -gz + \bar{P}$. We examine this relatively simple case because our previous work \citep{Zuccoli2025} did not cover depth-dependent base flows for which there can be a vertical shear at the free surface.

The first base velocity profile we consider is 
\begin{equation}
    U_0(y, z) = \mathcal{U}\left(1 + \frac{z}{\hinfty} \right) \e^{(1/\lambda - 1/\hinfty)z} \, \sin\Big(\frac{y}{\lambda}\Big), \quad -\hinfty\le z \le 0, \quad 0\le y\le 2\pi\lambda.
    \label{eq:base_test1_dimensional}
\end{equation}
where the bottom boundary must satisfy $\hinfty \ge \lambda$. The maximum free-surface velocity is $\mathcal{U}$. The parameter $\lambda$ is a length scale that controls the magnitude of the shear at the free surface. The maximum magnitudes of the vertical shear, $\partial_z U_0$, and spanwise shear, $\partial_y U_0$, on the free surface are both $\mathcal{U}/\lambda$. For simplicity, we choose to use the same length scale in both directions as rather than introduce a separate length scale for the $y$ direction. This is sufficient for our purposes here, and the example could readily be generalized to allow for different length scales in the two directions.

We non-dimensionalize velocities by $\mathcal{U}$, lengths by $\lambda$, and times by $\sqrt{\frac{\lambda}{g}}$,
so that $U_0 = \mathcal{U} \tilde{U}_0$, $(x,y,z) = (\lambda \tilde{x},\lambda \tilde{y},\lambda \tilde{z})$, $t = \sqrt{\frac{\lambda}{g}} \tilde{t}$, with tildes denoting non-dimensional variables. We then drop the tildes for clarity so that the non-dimensional velocity profile becomes
\begin{equation}
U_0(y, z) =  \underbrace{\left(1 + \frac{z}{h_\infty} \right) e^{(1 - 1/h_\infty)z}}_{\Zbase(z)} \, \underbrace{\sin(y)}_{\Ybase(y)}, \quad -h_\infty\le z \le 0, \quad 0\le y\le 2\pi,
\label{eq:base_test1}
\end{equation}
where $h_\infty = \hinfty/\lambda$ in the nondimensional depth and where $\Zbase(z)$ and $\Ybase(y)$ are functions of $z$ and $y$ defined as indicated. 
Because we have used an advective velocity scale but a gravitational time scale, the non-dimensionalization of the governing equations, either the full-depth equations \eqref{equ:linearised_euler} or the surface-only equations \eqref{equ:final_reduced_system} , results in 
the Froude number $\Fr = \mathcal{U}/\sqrt{g\lambda}$ appearing in the usual way as a factor in front of advective terms. In the following, we refer to equations \eqref{equ:linearised_euler} and \eqref{equ:final_reduced_system}, but implicitly understand these to mean the non-dimensional forms of these equations.

\subsection{Equations and numerical methods}\label{sec:eqns_and_num_methods_validation1}

The full-depth equations \eqref{equ:linearised_euler} can be then be entirely expressed in terms of the pressure. This is achieved by taking the divergence of the momentum equation \eqref{equ:horiz_mom_eq}, using the continuity equation \eqref{equ:cont_eq}, applying the convective derivative $D_t$ to the equation thus obtained, and finally exploiting the momentum equations for the $y$ and $z$ components, respectively. This yields
\begin{subequations}
    \begin{align} &
        \Big(\frac{\partial}{\partial t} + \Fr \Zbase(z)\Ybase(y)\frac{\partial}{\partial x}\Big)\nabla^2 p
        -2 \Fr \Zbase(z)\Ybase'(y)\frac{\partial^2 p}{\partial x\partial y} 
        -2 \Fr \Zbase'(z)\Ybase(y)\frac{\partial^2 p}{\partial x\partial z} = 0, \\ &
        \frac{\partial p}{\partial z} + 
        \Big(\frac{\partial}{\partial t} + \Fr \Ybase(y)\frac{\partial}{\partial x}\Big)^2 p = 0, \quad \mathrm{on} \quad z = 0, \\ &
        \frac{\partial p}{\partial z} = 0, \quad \mathrm{on} \quad z = -h_\infty,
    \end{align}%
    \label{equ:exact_perturbation_pb_test2_pressure_dimensionless}%
\end{subequations}%
where here prime symbol denotes differentiation with respect to a single variable (either $y$ or $z$) and $\nabla^2 = \partial^2_x + \partial^2_y + \partial^2_z$ is the three-dimensional Laplacian.

Seeking normal modes solutions of the form $p = \phi(y, z)e^{-\I\omega t + \I k x}$, with $k$ given, equations \eqref{equ:exact_perturbation_pb_test2_pressure_dimensionless} become
\begin{subequations}
    \begin{align} &
        \Big(k \Fr \,\Zbase(z)\Ybase(y) - \omega\Big)  \Big(\frac{\partial^2 \phi}{\partial y^2} + \frac{\partial^2\phi}{\partial z^2} - k^2\phi\Big) \nonumber \\
        & \qquad -2k \Fr \,\Zbase(z)\Ybase'(y)\frac{\partial\phi}{\partial y} 
        -2k \Fr \,\Zbase'(z)\Ybase(y)\frac{\partial \phi}{\partial z} = 0, \\ &
        \frac{\partial \phi}{\partial z} -\omega^2\phi + 2\omega k \Fr \Ybase(y)\phi - k^2 \Fr^2q^{2}_0(y)\phi = 0, \quad \mathrm{on} \quad z = 0, \\ &
        \frac{\partial \phi}{\partial z} = 0, \quad \mathrm{on} \quad z = -h_\infty,
    \end{align}%
    \label{equ:complete_normal_modes_pb2}%
\end{subequations}%
subject to periodic boundary conditions in $y$. System \eqref{equ:complete_normal_modes_pb2}
gives rise to a polynomial eigenvalue problem in the eigenvalues $\omega$ and eigenfunctions $\phi(y, z)$. From $\phi(y, z)$ the perturbation $h$ can be obtained via the dynamic boundary condition $h(y) = \phi(y, 0)$.

Turning now to the surface-only model equations \eqref{equ:final_reduced_system}, these are manipulated as follows. We take the horizontal divergence $\nabla_{\perp}\cdot$ of the horizontal momentum equations \eqref{equ:model_horiz_momentum_eqs} and exploit \eqref{equ:model_cont_eq} to get
\begin{equation}
    -\hat{D}^{2}_t\Big(\left.{\frac{\partial p}{\partial z}}\right|_{H_{0}}\Big) + \nabla^{2}_{\perp}h + 2\Fr f_{0}(0)q'_{0}(y)\frac{\partial v}{\partial x} + \Fr f'_{0}(0)q_{0}(y)\frac{\partial w}{\partial x} = 0.
\end{equation}
Now, we use vertical momentum equation \eqref{equ:model_vert_momentum_eq} and obtain
\begin{equation}
    \hat{D}^{3}_t w + \nabla^{2}_{\perp}h + 2\Fr f_{0}(0)q'_{0}(y)\frac{\partial v}{\partial x} + \Fr f'_{0}(0)q_{0}(y)\frac{\partial w}{\partial x} = 0.
\end{equation}
Using the kinematic boundary condition \eqref{equ:model_kin_BC},
\begin{equation}
    \hat{D}^{4}_t h + \nabla^{2}_{\perp}h + 2\Fr f_{0}(0)q'_{0}(y)\frac{\partial v}{\partial x} + \Fr f'_{0}(0)q_{0}(y)\frac{\partial }{\partial x} \hat{D}_t h = 0.
\end{equation}
Finally, applying $\hat{D}_t$ to this equation, and exploiting \eqref{equ:model_horiz_momentum_eqs} to eliminate $v$, yields the following PDE in the free surface height $h(x, y, t)$
\begin{equation}
    \hat{D}^{5}_t h + \hat{D}_t\nabla^{2}_{\perp}h + \Fr f'_{0}(0)q_{0}(y)\frac{\partial }{\partial x} \hat{D}^{2}_t h - 2\Fr f_{0}(0)q'_{0}(y)\frac{\partial^2 h}{\partial x \partial y} = 0. \label{equ:model_eq_in_h}
\end{equation}

For the base flow under consideration, the convective derivative acting on the free surface is $\hat{D}_{t} = \partial_t +  \Fr \, \Ybase(y)\partial_x$. Decomposing $h(x,y,t) \rightarrow h(y)e^{-\I\omega t + \I k x}$ into normal modes, equation \eqref{equ:model_eq_in_h} yields  the following polynomial eigenvalue problem
\begin{multline}
    \Big(k \Fr \, \Ybase(y) - \omega\Big)^{5}h + \Big(k \Fr \, \Ybase(y) - \omega\Big)(h'' - k^2 h)
    \\
    - k\Fr \, \Zbase'(0) \Ybase(y)\Big(k \Fr \Ybase(y)-\omega\Big)^2 h
    - 2k \Fr \, \Ybase'(y)h' = 0,
    \label{equ:model_eq_in_h_eigenvalue_Pb}
\end{multline}
subject to periodic boundary conditions in $y$. The term containing $\Zbase'$ is of particular interest because it accounts for the effect of the base vertical shear on the eigenvalue problem. 

We solve equations \eqref{equ:complete_normal_modes_pb2} and \eqref{equ:model_eq_in_h_eigenvalue_Pb} numerically by means of a spectral collocation method \citep{trefethen2000}, using Fourier modes along the spanwise direction $y$, and for~\eqref{equ:complete_normal_modes_pb2}, Chebyshev modes along the vertical direction $z$. This involves, for the numerical solution of the exact stability problem only, re-mapping the vertical domain from $(-h_\infty, 0)\rightarrow (-1, 1)$ when the depth is finite, and re-mapping $(-\infty, 0)\rightarrow (-1, 1)$ when the depth is infinite. For the latter, we employ the following map 
\citep{BoydSpectral2001}

\begin{equation}
    z = B_{\mathrm{map}}\tan\Big[\frac{\pi}{4}(z_c - 1)\Big], \quad \mathrm{with}\quad z \in (-\infty, 0], \quad \mathrm{and} \quad z_c \in [-1, 1],
    \label{equ:map_tan_infinite_domain}
\end{equation}
where $B_{\mathrm{map}}$ is a tuning parameter which sets the clustering of points close to the free surface $z = 0$. We use spatial discretizations $(N_y, N_z) = (52, 100)$ along the spanwise and vertical directions respectively, and use $B_{\mathrm{map}} = 4$. 
Further details on the numerical discretization and convergence of the eigensolutions for this first validation problem are contained in appendix~\ref{sec:appendix_numerics}.
It is worth appreciating how faster the computations using the surface-only model equations are with respect to the full-depth counterpart. Estimates of the mean computational times give 
\(\ T^{(N)}_{\mathrm{mean}}=6\times 10^{3}\,\mathrm{s}\) for the full-depth Euler equations, while \(\ T^{(M)}_{\mathrm{mean}}=8\times 10^{-2}\,\mathrm{s}\) for the surface-model equations. It follows that the model is orders of magnitude faster than the full-depth equations. 

Additionally, when solving the surface-model equation \eqref{equ:model_eq_in_h_eigenvalue_Pb}, its spectrum comprises in general both ``physical" and ``unphysical" eigenvalues. This feature arises because the model equations intrinsically cannot distinguish whether the fluid extends either infinitely deep below or infinitely deep above the base free surface. Since our goal is to compute modes which only decay vertically, in order to remove the unphysical eigenvalues we define and measure some sort of average axial decay rate. If the decaying of a mode is sufficiently large, then that mode is retained. If not, that is discarded. Further details are contained in appendix \ref{sec:unphysical_modes_model}.

\subsection{Results}

We now compare eigenvalues and eigenmodes from the surface-only model \eqref{equ:model_eq_in_h_eigenvalue_Pb} to those from the full-depth equations \eqref{equ:complete_normal_modes_pb2}. We also assess directly whether the closure condition \eqref{equ:closure_final} is satisfied by the full-depth solutions. Throughout this section we fix the streamwise wavenumber at $k = 1$, corresponding to eigenmodes whose streamwise wavelength is comparable to the length scales of the base flow. 
 
\begin{figure}
    \centering
    \includegraphics[width=0.6\textwidth]{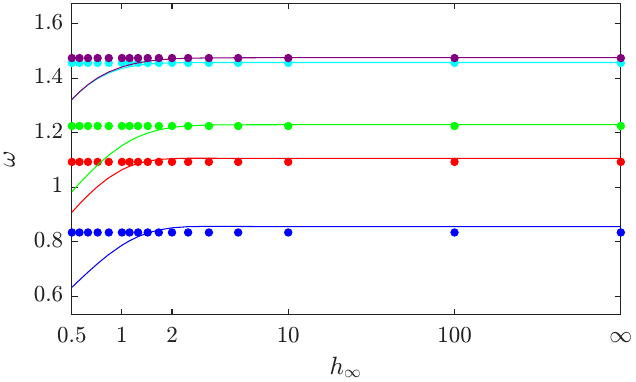}
    \caption{Trend of the first five positive eigenvalues as function of dimensionless depth $h_\infty$ for $\Fr = 0.5$. The base flow velocity profile is given by equation \eqref{eq:base_test1}.
    Eigenvalues from full-depth equations, viewed as exact, are shown in full lines. Eigenvalues from the surface-only model are shown as points.}
    \label{fig:evals_Vs_eps_test1}
\end{figure}%
In figure~\ref{fig:evals_Vs_eps_test1} we show the first five positive eigenvalues as function of $h_\infty$ from the full-depth (full lines) and surface-only (points) equations. The Froude number is $\Fr = 0.5$. 
All eigenvalues are real. We plot only the positive half of the spectrum, since the spectrum is symmetric about zero (discussed further below).
It can be seen that the surface-only eigenvalues not only accurately match those from the exact numerical calculations in the deep-water limit ($h_\infty\rightarrow\infty$), but also for $h_\infty$ of order unity. Significant deviations between the surface-only and full-depth cases begin at $h_\infty$ between 1 and 2. As the fluid depth becomes shallow, $h_\infty < 1$, the deviations become large. This is in close agreement with the results obtained in \citet{Zuccoli2025} for an axisymmetric flow. 
\begin{figure}
    \centering
    \includegraphics[width=6.3cm, height=3.8cm]{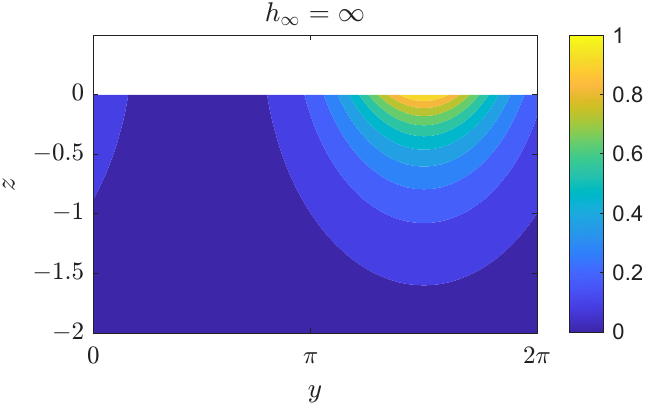}
    \hspace{0.5cm}
    \includegraphics[width=5.5cm, height=3.8cm]{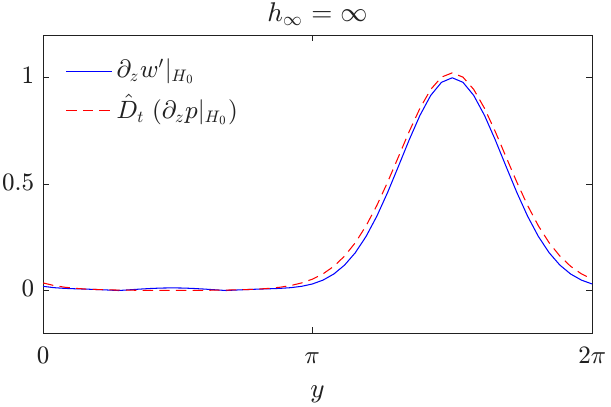}
    \par
    \vspace{0.5cm}
    \includegraphics[width=6.3cm, height=3.8cm]{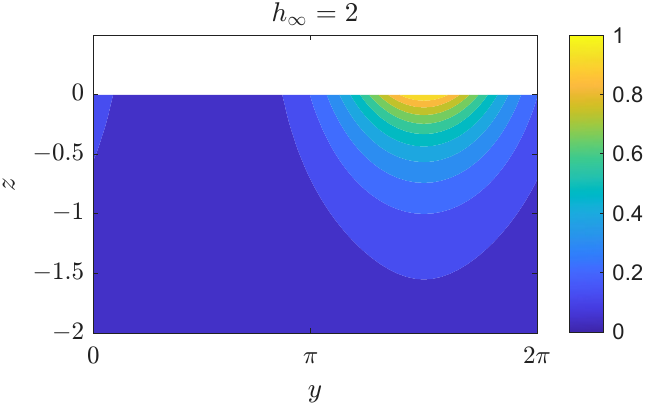}
    \hspace{0.5cm}
    \includegraphics[width=5.5cm, height=3.8cm]{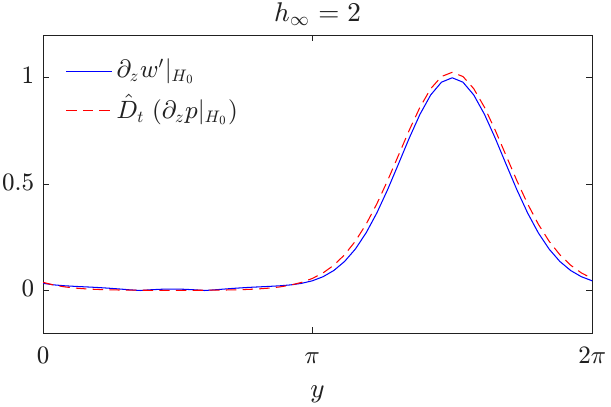}
    \par
    \vspace{0.5cm}
    \includegraphics[width=6.3cm, height=3.8cm]{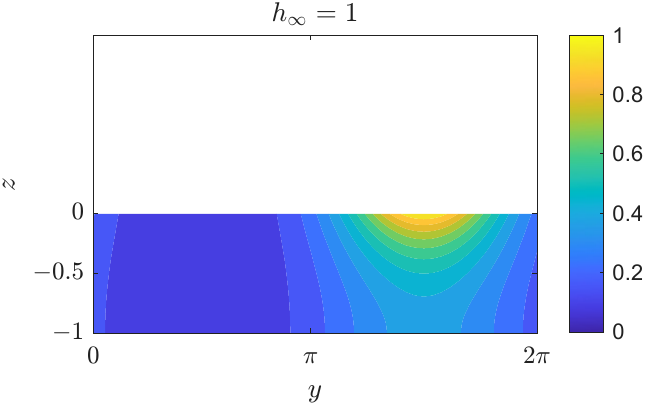}
    \hspace{0.5cm}
    \includegraphics[width=5.5cm, height=3.8cm]{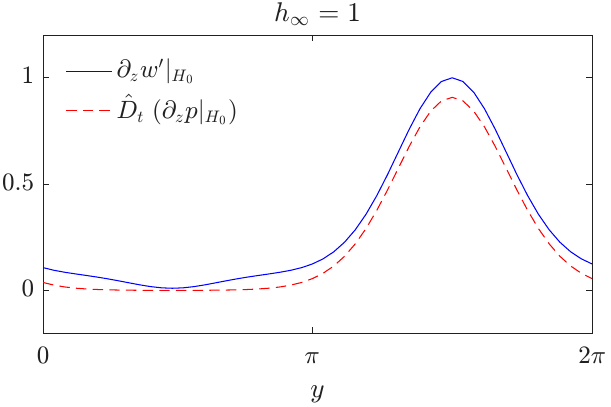}
    \par
    \vspace{0.5cm}
    \includegraphics[width=6.3cm, height=3.8cm]{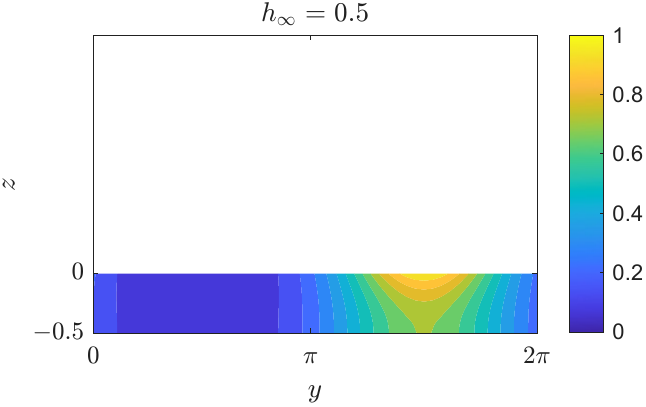}
    \hspace{0.5cm}
    \includegraphics[width=5.5cm, height=3.8cm]{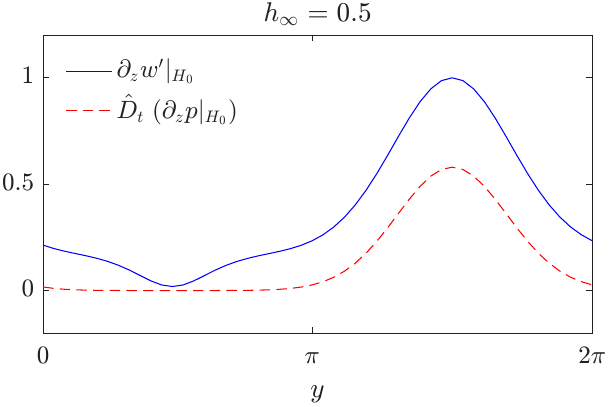}
    \caption{Accuracy of the closure condition \eqref{equ:closure_final} for the first surface-wave modes with Froude number $\Fr = 0.5$ at four values of dimensionless fluid depth $h_{\infty}$. Left panels illustrate the modes with contours of the pressure field $\phi(y, z)$. The modulus of both left-hand and right-hand sides of the closure condition are evaluated and plotted in the right panels, with the eigenfunction normalized so that the maximum value of the left-hand side is 1.}
    \label{fig:surface_plots_TEST1}
\end{figure}%

To gain further insight into the effect of fluid depth, we plot in figure~\ref{fig:surface_plots_TEST1} (left panels) pressure contours of the first eigenfunction from the full-depth computations at four values of dimensionless fluid depth $h_{\infty} = 0.5, 1, 2, \infty$. The eigenmodes at $h_\infty = 1$ and $h_\infty = 0.5$ are clearly influenced by the bottom boundary, and unsurprisingly, these cases are not in the deep-water regime and the model is not expected to hold.  

Additionally, we assess directly whether the closure condition \eqref{equ:closure_final} holds for the exact full-depth eigenmodes. For the case under consideration, the dimensionless form of the closure equation reduces to
\begin{equation}
\left.{\frac{\partial \wpert}{\partial z}}\right|_{H_{0}} = 
\hat{D}_{t}\Big(\left.\frac{\partial p}{\partial z}\right|_{H_{0}}\Big).
\end{equation}
where now $\hat{D}_t = -\I\omega + \I k q_0(y)$ from the normal modes representation. The left-hand and right-hand sides of this equation are plotted in figure~\ref{fig:surface_plots_TEST1} (right panels). At $h_\infty = \infty$ and $h_\infty = 2$ there is close agreement between the curves, implying that the model is accurately closed in these cases. The agreement is not perfect, however, even at $h_\infty = \infty$. For $h_\infty = 1$ and $h_\infty = 0.5$ the closure condition is clearly failing and hence one could not expect the model to produce reliable results. 

\begin{figure}
    \centering
    \includegraphics[width=4.6cm]{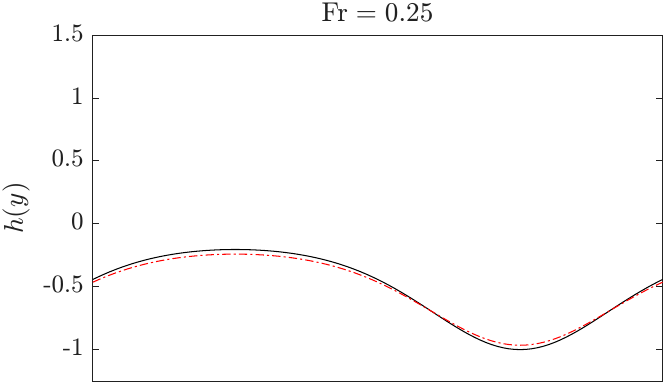}
    \hspace{0.2cm}
    \includegraphics[width=4cm]{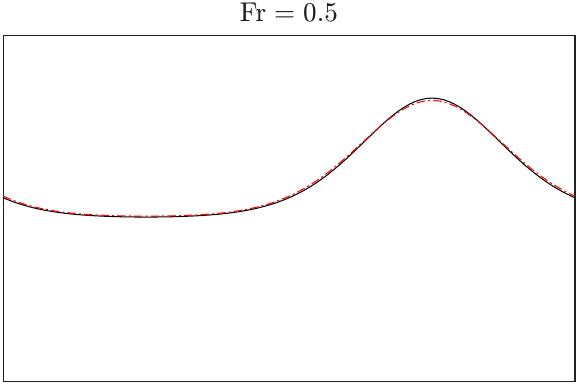}
    \hspace{0.2cm}  
    \includegraphics[width=4cm]{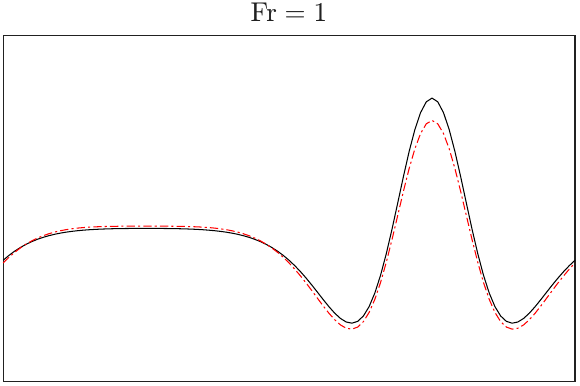}
    \par
    \vspace{0.5cm}
    \includegraphics[width=4.6cm]{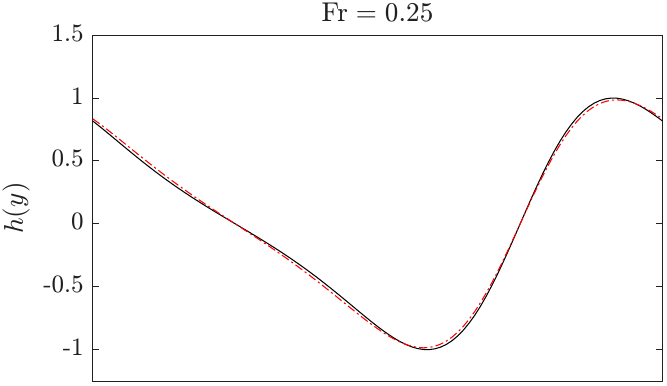}
    \hspace{0.2cm}
    \includegraphics[width=4cm]{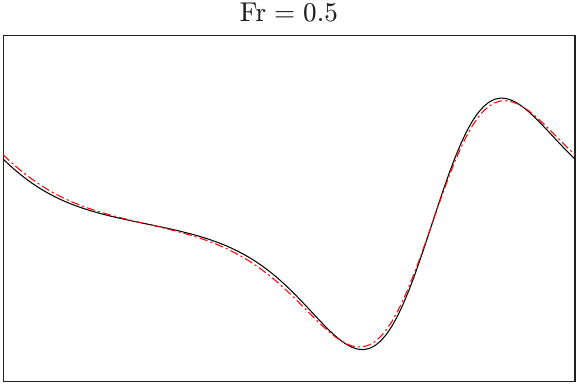}
    \hspace{0.2cm}
    \includegraphics[width=4cm]{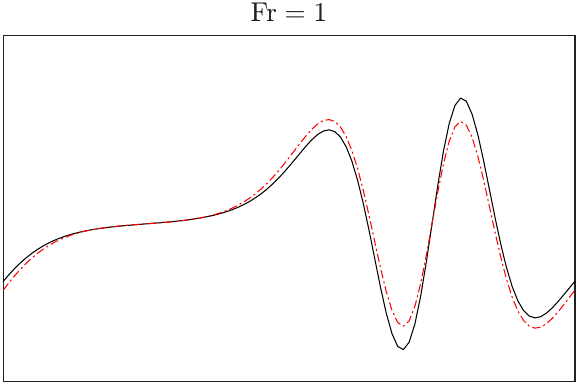}
    \par
    \vspace{0.5cm}
    \includegraphics[width=4.6cm]{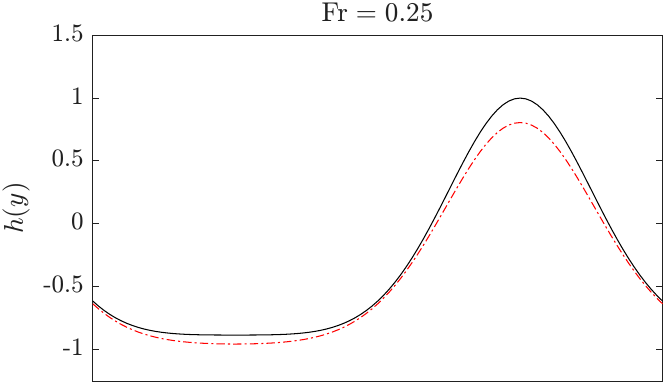}
    \hspace{0.2cm}
    \includegraphics[width=4cm]{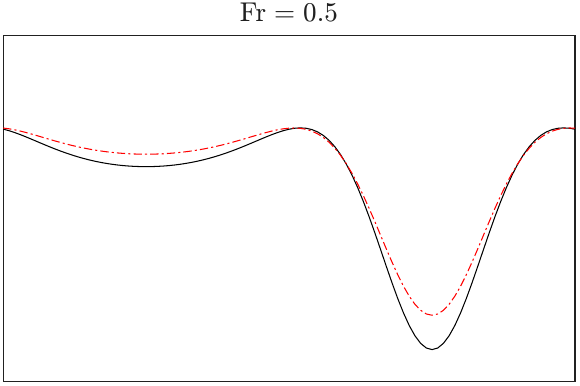}
    \hspace{0.2cm}
    \includegraphics[width=4cm]{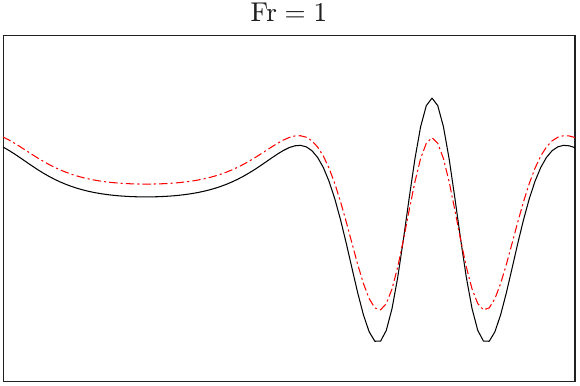}
    \par
    \vspace{0.5cm}
    \includegraphics[width=4.6cm]{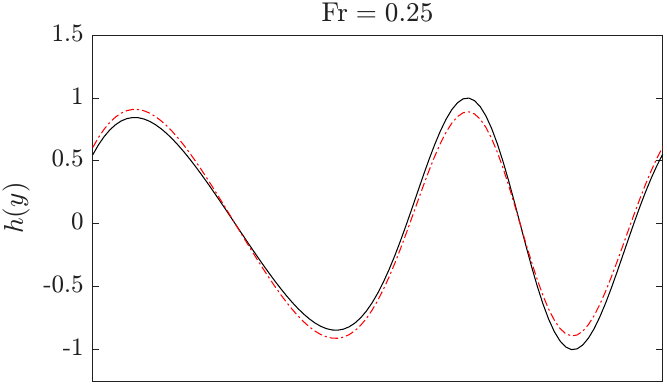}
    \hspace{0.2cm}
    \includegraphics[width=4cm]{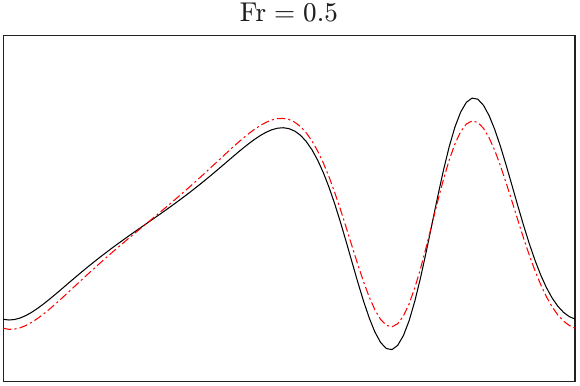}
    \hspace{0.2cm}
    \includegraphics[width=4cm]{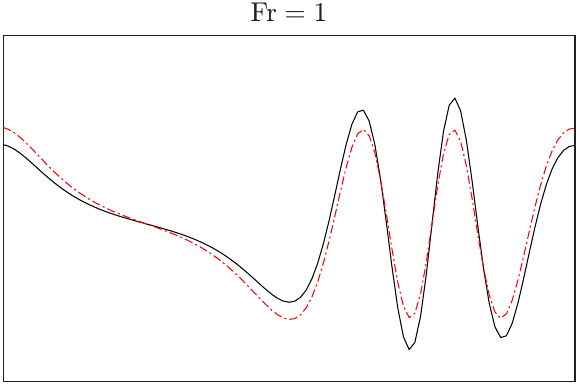}
    \par
    \vspace{0.5cm}
    \includegraphics[width=4.6cm]{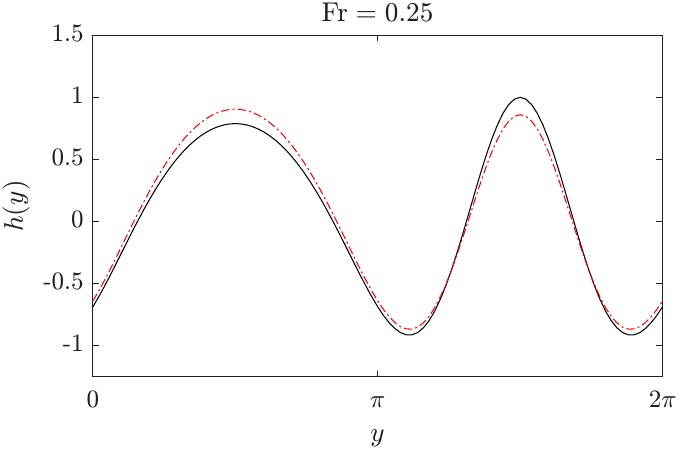}
    \hspace{0.2cm}
    \includegraphics[width=4.0cm]{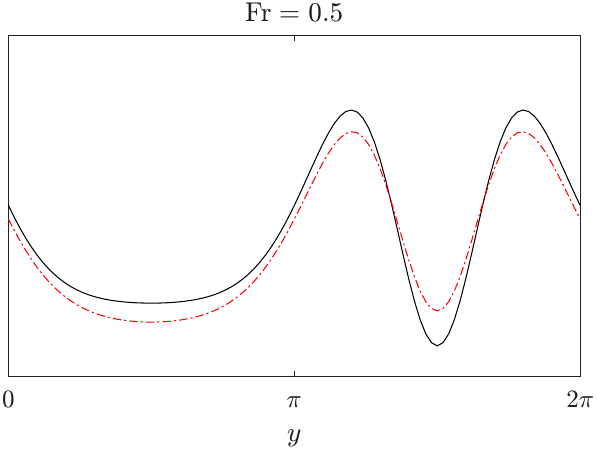}
    \hspace{0.2cm}
    \includegraphics[width=4.0cm]{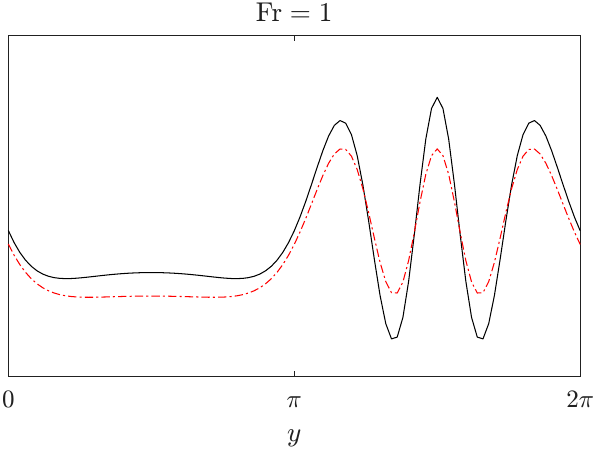}
    \caption{Comparison of the surface-only and full-depth eigenfunctions for Froude numbers $\Fr = 0.25, 0.5, 1$, with $h_\infty = \infty$. The free-surface elevation $h(y)$ is plotted for the first five modes, corresponding to the five eigenvalues seen in figure~\ref{fig:evals_Vs_eps_test1}. The panels are ordered by frequency with the top panel the first (lowest-frequency) mode. full-depth modes are shown in full black lines and surface-only modes dashed red lines.
    }
    \label{fig:comparison_eigenfunctions_numerics_Vs_model_test1}
\end{figure}%

We continue by comparing the surface-only eigenmodes to the eigenmodes from the full-depth equations. Figure~\ref{fig:comparison_eigenfunctions_numerics_Vs_model_test1} plots the first five eigenmodes in terms of the free-surface elevation $h$ for $\Fr = 0.25, 0.5, 1$, with $h_\infty = \infty$. All eigenmodes are normalized to have an $L_2$ norm of $1$, the normalization convention used throughout this work. While differences are visible between the two sets of modes, overall the surface-only eigenmodes capture very well the corresponding eigenmodes from the full-depth equations.
It is evident that the eigenmodes are not symmetric under reflection about $y=\pi$ or, equivalently, taking into account the periodic domain, about $y=0$. It is straightforward to show from the symmetry of the base flow that if $h(y)$ is an eigenmode with real frequency $\omega$, then $h(-y)$ is also an eigenmode with frequency $-\omega$. Hence, the negative half of the spectrum, not shown in figure~\ref{fig:evals_Vs_eps_test1}, corresponds to the reflected versions of the modes shown in figure~\ref{fig:comparison_eigenfunctions_numerics_Vs_model_test1}.
Since the eigenvalues are real, the time-dependent modes can be interpreted as travelling waves in the streamwise direction of the form $h(y)\cos(kx-\omega t)$. The modes shown in figure~\ref{fig:comparison_eigenfunctions_numerics_Vs_model_test1} corresponds to waves travelling in the $+\vect{\hat{x}}$ direction, since $k=1$, $\omega>0$.

\begin{figure}
    \centering
    \includegraphics[width=0.6\textwidth]{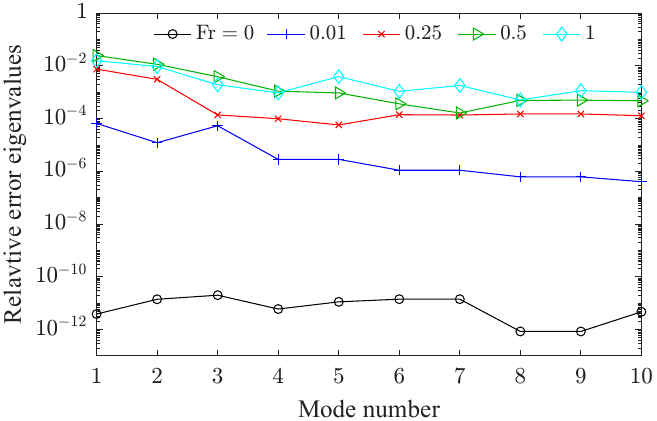}
    \caption{Relative error of the first ten eigenvalues between full-depth and surface-only model computations for different Froude numbers.
    }
    \label{fig:realErr_evals_different_Froude}
\end{figure}

Finally, we address the quantitative accuracy of the surface-only eigenvalues compared with those from the full-depth equations. 
Let the full-depth eigenvalues be given by $\omega^{(FD)}_i$ and the surface-only eigenvalues $\omega^{(SO)}_i$. In figure \ref{fig:realErr_evals_different_Froude} we plot the relative error between the two $|\omega^{(FD)}_{i} - \omega^{(SO)}_{i}|/|\omega^{(FD)}_{i}|$ for the first ten eigenvalues computed at five values of Froude number $\Fr = [0, 0.01, 0.25, 0.5, 1]$. The error is below $2 \%$ over the entire range of modes considered, confirming the accuracy and the robustness of the surface-only model in these cases. 
For \(\Fr=0\), the agreement is at the level of numerical precision because the base flow is motionless, and hence both the full stability equations and the surface-only model admit the same solutions, with exact eigenvalues given by $\omega_n = \sqrt[4]{n^2 + k^2}$, $n \in \mathbb{Z}$ (see appendix~\ref{sec:app_exact_solutions_flat_FS} for further details).

\subsection{Velocity profile with depth-dependent surface shear}

For the base velocity profile studied thus far, the surface-velocity gradients $\partial_z U_0$ and $\partial_y U_0$ are independent of the fluid depth. This is why the surface-only eigenvalues in figure~\ref{fig:evals_Vs_eps_test1} do not vary with depth $h_\infty$; the surface-only model contains the shear of the base flow at the free surface and not fluid depth directly. 
However, it is also natural for the vertical shear of a fluid flow to increase as the fluid depth decreases and so we briefly consider such a case using a variant of the base velocity profile studied thus far. Rather than using $\Zbase(z)$ appearing \eqref{eq:base_test1}, we set
\begin{equation}
\Zbase(z) = \frac{\sinh(1+z/h_\infty)}{\sinh(1)}.
\label{eq:profile2}
\end{equation}
The base flow is otherwise unchanged from the previous test case. For the surface-only eigenvalue problem \eqref{equ:model_eq_in_h_eigenvalue_Pb}, the only change 
is in $\Zbase'$ at the free surface. Previously this was $\Zbase'(0) = 1$, while now it is $\Zbase'(0) = \coth(1)/h_\infty$. 

\begin{figure}
    \centering
    \includegraphics[width=0.6\textwidth]{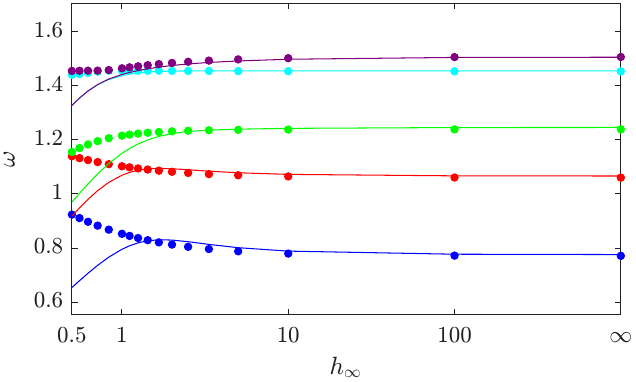}
    \caption{Trend of the first five positive eigenvalues as function of dimensionless depth $h_\infty$ for $\Fr = 0.5$. The base flow velocity profile is given by equation \eqref{eq:profile2}.
    Eigenvalues from full-depth equations, viewed as exact, are shown in full lines. Eigenvalues from the surface-only model are shown as points.}
    \label{fig:evals_Vs_eps_test2}
\end{figure}

In figure~\ref{fig:evals_Vs_eps_test2} we show the first five positive eigenvalues as function of $h_\infty$ from the full-depth equations (full lines) and the surface-only model (points). The Froude number is $\Fr = 0.5$. Comparing figure~\ref{fig:evals_Vs_eps_test2} with figure~\ref{fig:evals_Vs_eps_test1}, we see that in both cases there is good agreement between the surface-only model and full-depth eigenvalues down to fluid depths $h_\infty$ of order unity. 
However, unlike in figure~\ref{fig:evals_Vs_eps_test1}, in figure~\ref{fig:evals_Vs_eps_test2} we see that eigenvalues begin to depend depth when $h_\infty$ is of order 10. This is due to the variation of the base shear with depth. The surface-only model clearly picks up the initial deviations of the eigenvalues from the infinite-depth limit as $h_\infty$ decreases. Once the fluid depth becomes shallow, $h_\infty < 1$, the surface-only model fails and the deviations become large.
While not shown, a direct validation of the closure condition for this profile leads to results very similar to what is shown in figure~\ref{fig:surface_plots_TEST1}. Likewise, a comparison of eigenfunctions strongly resembles what is seen in figure~\ref{fig:comparison_eigenfunctions_numerics_Vs_model_test1}.

\section{Validation 2 --- waves on a Gerstner wave}\label{sec:validation_model2}

Gerstner waves \citep{gerstner1809theorie, RankineGerstnerWaves}, also known as trochoidal waves, are exact solutions to the Euler equations with both depth-dependent velocity and free-surface deformation, and as such these waves provide an interesting and important class of base flows to test our reduced model. Gerstner-waves are given in Lagrangian coordinates and are valid in the infinitely deep-water regime. When considered in a reference frame moving with the wave speed, these provide exact steady base flows with nonzero vorticity distribution. In the following, we compute linear surface waves propagating on Gerstner waves in our surface-only model and compare them with corresponding surface waves from full-depth Euler calculations.

\subsection{The base flow}

Let $(x, z)$ be Eulerian spatial coordinates in a frame moving with the wave, let $(\mathcal{X}, \mathcal{Z})$ be Lagrangian coordinate in a fixed reference frame, and let $(a, b)$ the Lagrangian coordinates in the reference frame moving with the wave, i.e. $a = \mathcal{X} + ct$, $b = \mathcal{Z}$, with $c = \sqrt{g/k}$ denoting the phase speed of the wave in deep-water. We define the following Euler-Lagrange map between the two coordinate systems
    \begin{align}
        x &= a + \frac{1}{k}e^{kb}\sin(ka), &
        z &= b - \frac{1}{k}e^{kb}\cos(ka),
        \label{equ:euler_lagr_map}
    \end{align}
with $k$ being the wavenumber of the Gerstner wave and $b \in (-\infty, \BS]$, where $b = \BS \le 0$ dictates the location of the steady free surface. The corresponding velocity components in the steady Lagrangian reference frame are given by
    \begin{align}
        U_0 &= c + ce^{kb}\cos(ka), &
        W_0 &= ce^{kb}\sin(ka).
        \label{equ:velocity_lagr}
    \end{align}
The base free surface $z = H_{0}(x)$ is determined implicitly by fixing $b=\BS$ and varying $a$ in the Euler-Lagrange map \eqref{equ:euler_lagr_map}.
In particular, for large nagative $\BS$ (in the limit $\BS\rightarrow -\infty$) the base velocity field becomes uniform in the horizontal direction and the free surface flat. 
For increasing values of $\BS$ the free surface deformation becomes more significant, with the limiting case of $\BS = 0$ giving a cusp-shaped free surface.

\begin{figure}
    \centering
    \includegraphics[width=0.6\textwidth]{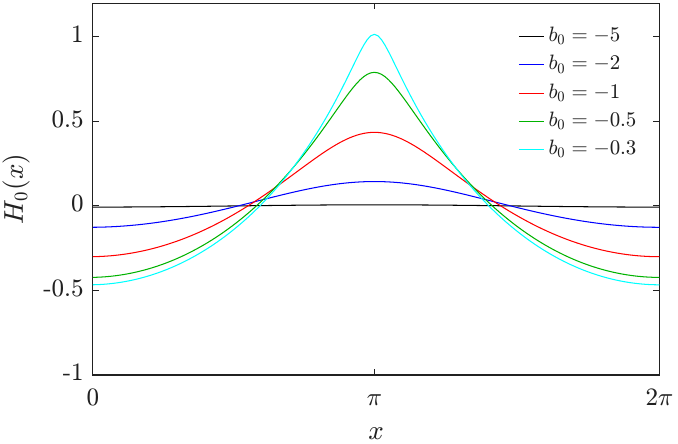}
    \caption{Base free surfaces $H_{0}(x)$ for representative values of $\BS$.}
    \label{fig:gerstner_base_FS}
\end{figure}

We non-dimensionalize lengths by $1/k$, so that $a = \tilde{a}/k$, $x = \tilde{x}/k$, etc.\ (with tildes denoting nondimensional variables), and times by $1/\sqrt{gk}$, so that $\tilde{c} = 1$.  We now drop the tildes for clarity.  Additionally, we shift the coordinate $z$ vertically upward by $-\BS$, so that the mean level of the free surface $H_0(x)$ is approximately at $z=0$ for any value of $\BS$. The dimensionless base flow is then given by
\begin{subequations}
    \begin{align} &
        x = a + e^{b}\sin(a),  \\ &
        z = b - \BS - e^{b}\cos(a), \label{eq:G_base_z} \\ &
        U_0 = 1 + e^{b}\cos(a) \\ &
        W_0 = e^{b}\sin(a),
    \end{align}%
\end{subequations}%
where $a \in [0, 2\pi]$ and $b \in [-\infty, \BS]$. The base flows are thus parameterized by the single parameter $\BS$, where $\BS < 0$.
Free-surface heights for representative values of $\BS$ are shown in figure~\ref{fig:gerstner_base_FS}.

\subsection{Perturbations dynamics}

The Gerstner base flows are two-dimensional, and we will only considered perturbations that are also two dimensional, with velocity field given by $\upertvec = (\upert, \wpert)$. 
Note that Gerstner waves are known to be unstable to three-dimensional perturbations for $\BS > -\ln 3$ \citep{leblanc2004local}, although this short-wavelength instability is absent in the two-dimensional case considered here (as can be seen by choosing a mode wave vector in the $x$-$z$ plane in equation~(3.3) in \citealp{leblanc2004local}).
We assume the perturbations to be characterized by the same spatial scale $1/k$ and time scale $1/\sqrt{gk}$ as for the base flow. The exact two-dimensional perturbations equations \eqref{equ:linearised_euler} in dimensionless form then read
\begin{subequations}\begin{align} &
\frac{\partial\upert}{\partial t} + \vect{U}_{0}\cdot\nabla \upert + \vect{\upertvec}\cdot\nabla U_{0} + \frac{\partial p}{\partial x} = 0, \label{equ:x_mom_gerstner} \\ &
\frac{\partial \wpert}{\partial t} + \vect{U}_{0}\cdot\nabla \wpert + \vect{\upertvec}\cdot\nabla W_{0} + \frac{\partial p}{\partial z} = 0, \label{equ:z_mom_gerstner} \\ &
\frac{\partial \upert}{\partial x} + \frac{\partial \wpert}{\partial z} = 0, \label{equ:cont_eq_gerstner} \\ &
\frac{\partial p}{\partial t} = g_{0}\Big(\wpert - \frac{\mathrm{d}H_{0}}{\mathrm{d}x}\upert\Big) - U_{0}
\Big[\frac{\partial p}{\partial x} + \frac{\mathrm{d}H_{0}}{\mathrm{d}x}\frac{\partial p}{\partial z} - \frac{1}{g_{0}}\frac{\mathrm{d}g_{0}}{\mathrm{d}x}p\Big] + \varphi_{0}p, \quad \mathrm{on} \quad z = H_{0}.
\end{align}\label{equ:full_linearised_euler_gerstner}%
\end{subequations}%
where we have substituted the dynamic boundary condition 
\eqref{equ:dyn_BC} into the kinematic boundary\eqref{equ:kin_BC}, to obtain a single boundary condition in $p$. 

The quantities $g_{0}(x)$ and $\varphi_{0}(x)$ are given by
\begin{equation}
        g_{0} \!= 1 + U_{0}\frac{\partial W_{0}}{\partial x} + W_{0}\frac{\partial W_{0}}{\partial z}, \quad \mathrm{and}\;\;  
        \varphi_{0} \!= \frac{\partial W_{0}}{\partial z} - \frac{\mathrm{d}H_{0}}{\mathrm{d}x}\frac{\partial U_{0}}{\partial z}, \quad \text{both on}\; z = H_{0}.
\end{equation}
Additionally, periodic boundary conditions are imposed in the $x-$direction, identifying $x=2\pi$ with $x=0$. One could consider the perturbation problem over multiple wavelengths of the base flow, but we do not pursue that here.

As the upper boundary of the domain is given by $H_{0}(x)$, which is not constant, it is numerically convenient to introduce the shifted vertical coordinate $\xi = z - H_{0}(x) \in (-\infty, 0]$.
The derivatives transform into $\partial_x \rightarrow \partial_x - (\mathrm{d}H_{0}/\mathrm{d}x)\partial_{\xi}$ and $\partial_z\rightarrow \partial_{\xi}$, so that equations \eqref{equ:full_linearised_euler_gerstner} are re-written accordingly in terms of $(x, \xi, t)$.
In order to discretize the equations along the vertical direction we use the same map \eqref{equ:map_tan_infinite_domain} as for the previous validation problem. Finally, we decompose the unknowns into normal modes of the form $[\upert, \wpert, p] = \Real\Big([\hat{u}(x, \xi), \hat{w}(x, \xi), \hat{p}(x, \xi)]e^{-\I\omega t}\big) = \Real\Big(\boldsymbol{\hat{w}}(x, \xi)e^{-\I\omega t}\Big)$. Note that, since $[\upert, \wpert, p]$ is real, if $(\boldsymbol{\hat{w}},\omega)$ is a solution then so too is $(\boldsymbol{\hat{w}}^*,-\omega^*)$, where $^*$ denotes the complex conjugate.  This yields the following generalized eigenvalue problem in the eigenvalues $\omega$
\begin{equation}
    \mathcal{A}(\partial_x, \partial_{\xi}; \BS)\boldsymbol{\hat{w}} = \omega\mathcal{B}(\BS)\boldsymbol{\hat{w}}.
    \label{equ:generalized_eval_PB_Gerstner}
\end{equation}
Problem \eqref{equ:generalized_eval_PB_Gerstner} is solved numerically in primitive variables using a collocation spectral method, similar to the one employed in section \ref{sec:validation_model1} and described in appendix~\ref{sec:appendix_numerics}.

For reference, the surface-only model equations for this case reduce to
\begin{subequations}\begin{align} &
\hat{D}_t u + 
u \left. \partial_x U_0\right|_{H_{0}} + 
w \left. \partial_z U_0\right|_{H_{0}}
+ \partial_x(g_0 h) - 
H_0' \left.{\frac{\partial p}{\partial z}}\right|_{H_{0}} = 0, \\ &
\hat{D}_t w +
u \left. \partial_x W_0\right|_{H_{0}} + 
w \left. \partial_z W_0\right|_{H_{0}}
+ \left.{\frac{\partial p}{\partial z}}\right|_{H_{0}}  = 0,  \\ &
\hat{D}_{t}\Big(\left.{\frac{1}{g_0}\frac{\partial p}{\partial z}}\right|_{H_{0}}\Big) - \frac{\varphi_0}{g_0}\left.{\frac{\partial p}{\partial z}}\right|_{H_{0}} + 
\partial_x u
= 0, \\ &
\hat{D}_t h
+ u H_0'
- \varphi_{0}h - w = 0, 
\end{align}%
\label{equ:reduced_system_Gerst}%
\end{subequations}
where $\hat{D}_t = \left( \partial_t + U_0|_{H_0}  \partial_x \right)$. This is a system of four PDEs in primitive variables $(u, w, h, \partial_z p|_{H_{0}})$. Seeking normal modes of the form $e^{-\I\omega t}$ results in an eigenvalue problem. The spatial domain of this eigenvalue problem is simply the periodic interval $x \in [0, 2\pi]$ and is readily solved numerically. 

Likewise for the numerical implementation of the first validation test described in section \ref{sec:eqns_and_num_methods_validation1}, even in this case the surface-model equations \eqref{equ:reduced_system_Gerst} give rise to some unphysical eigenvalues which need to be discarded accordingly. Details can be found in appendix \ref{sec:unphysical_modes_model}.

\subsection{Results}

\begin{figure}
    \centering
    \includegraphics[width=9cm, height=6cm]{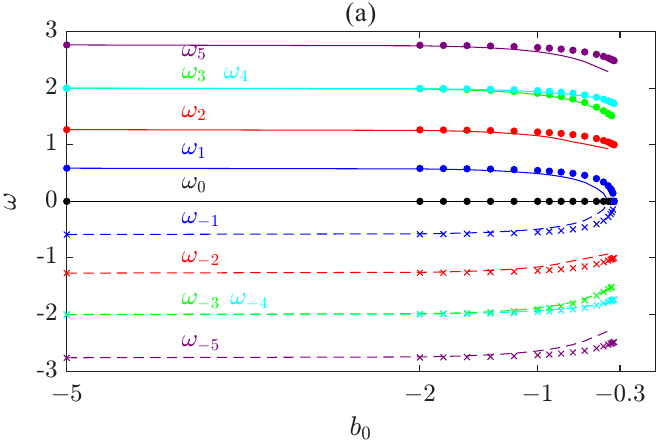}
    \hspace{0.5cm}
    \includegraphics[width=3.75cm, height=6cm]{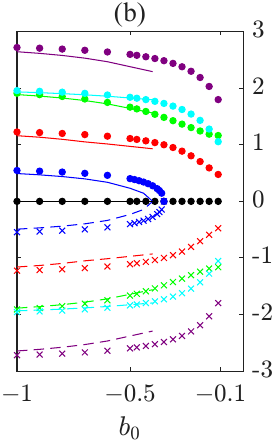}
    \caption{
    (a) Comparison of the eigenvalues from the full-depth Euler stability problem (solid lines) and the surface-only model equations (markers) up to $\BS = -0.3$. (b) Continuation of the eigenvalue branches for the surface-model equations alone for values of $\BS$ up to $-0.1$, where the peak of the base free surface is close to forming a cusp.
    }
    \label{fig:comparison_evals_num_vs_model}
\end{figure}

We compare eigensolutions obtained from the full-depth Euler equations \eqref{equ:full_linearised_euler_gerstner} with those coming from the surface-only model equations \eqref{equ:reduced_system_Gerst} for various free-surface deformations of the base solution, corresponding to different values of $\BS$. 
We begin with a comparison of eigenvalues shown in figure~\ref{fig:comparison_evals_num_vs_model}(a). 
All eigenvalues are real, corresponding to stable modes. 
We index the eigenvalues such that $\cdots < \omega_{-1} < \omega_0 = 0 < \omega_1 < \cdots$.
For $\BS = -5$, the peak-to-peak variation in the free-surface elevation of the base flow is $O(10^{-2})$, which is already close to the flat case corresponding to the limit $\BS \rightarrow -\infty$.
Accordingly, each eigenvalue $\omega_j$ tends to a closed-form solution valid in this limit, 
$\omega_j = n\pm\sqrt{|n|}$ for some $n \in \mathbb{Z}\setminus\{0\}$
(see appendix~\ref{sec:appendix_gerstner}).
As $\BS$ is increased, corresponding to larger wave amplitude of the base solution, the surface-only eigenvalues follow closely the full-depth eigenvalues up to $\BS \simeq -1$. Beyond this point, the discrepancy between the two eigenvalue branches becomes more noticeable, although the surface-only model qualitatively captures the behaviour of the full-depth Euler eigenvalues. In particular, it is worth noting that the first non-zero eigenvalues,
$\omega_{\pm 1}$, collapse to zero at a finite value $\BS = -0.43$ for the full-depth Euler system and $\BS = -0.35$ for the surface-only model. The eigenvalues then split into pairs of complex-conjugate eigenvalues with zero real part, resulting in an instability of the system described in \S \ref{sec:onset_instability}.

\begin{figure}
    \centering
    \includegraphics[width=4.6cm, height=2.6cm]{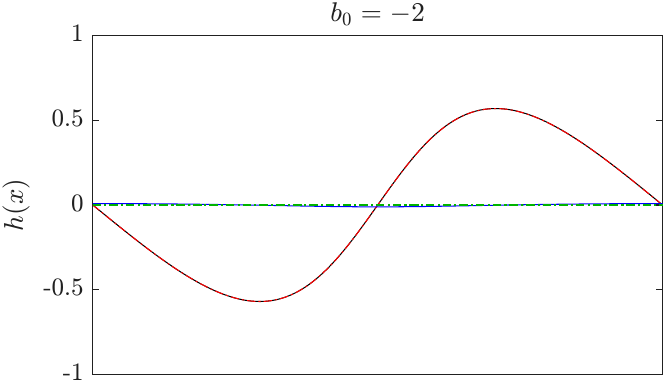}
    \hspace{0.2cm}
    \includegraphics[width=4cm, height=2.6cm]{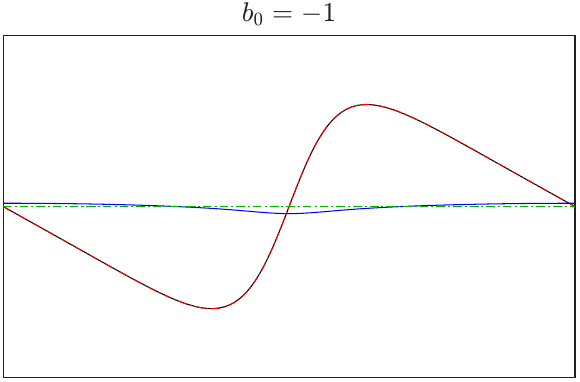}
    \hspace{0.2cm}
    \includegraphics[width=4cm, height=2.6cm]{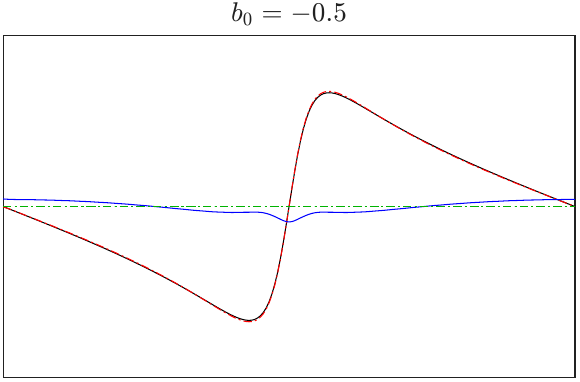}
    \par
    \vspace{0.5cm}
    \includegraphics[width=4.6cm]{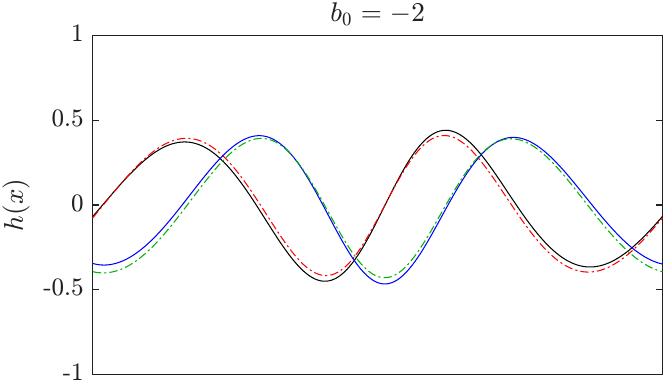}
    \hspace{0.2cm}
    \includegraphics[width=4cm]{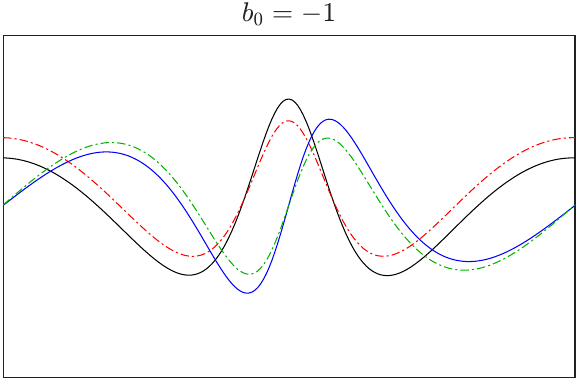}
    \hspace{0.2cm}
    \includegraphics[width=4cm]{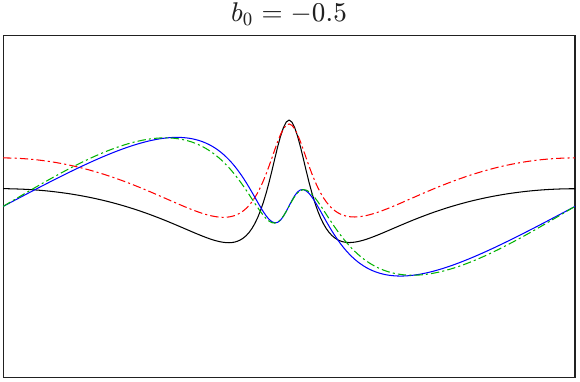}
    \par
    \vspace{0.5cm}
    \includegraphics[width=4.6cm]{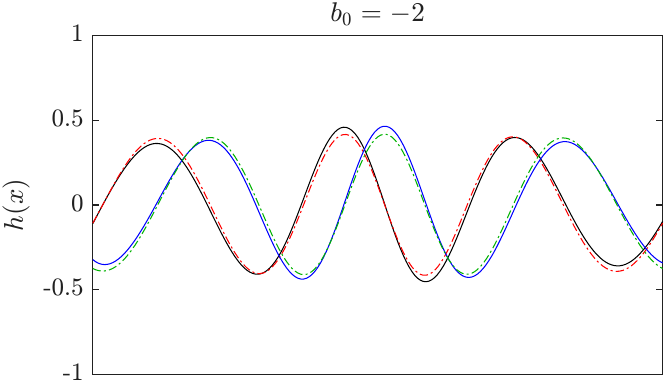}
    \hspace{0.2cm}
    \includegraphics[width=4cm]{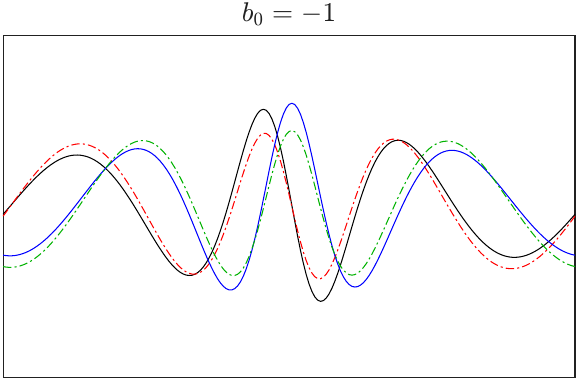}
    \hspace{0.2cm}
    \includegraphics[width=4cm]{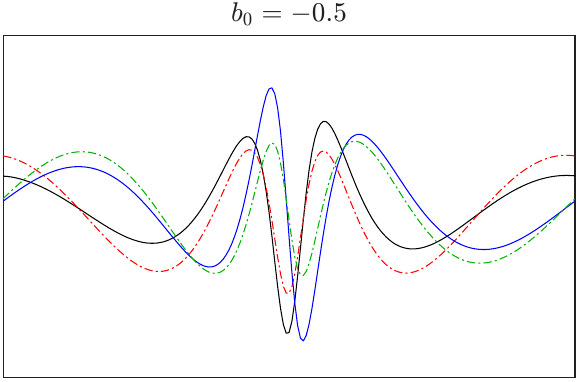}
    \par
    \vspace{0.5cm}
    \includegraphics[width=4.6cm]{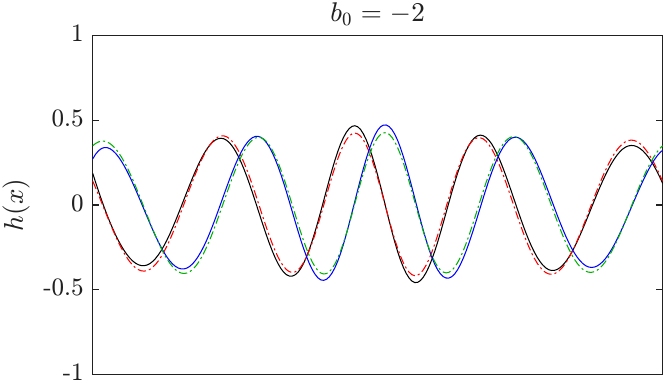}
    \hspace{0.2cm}
    \includegraphics[width=4cm]{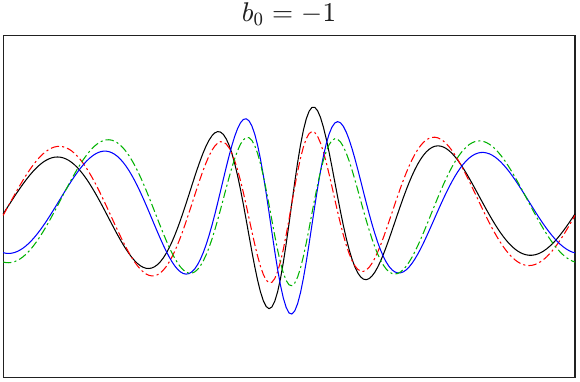}
    \hspace{0.2cm}
    \includegraphics[width=4cm]{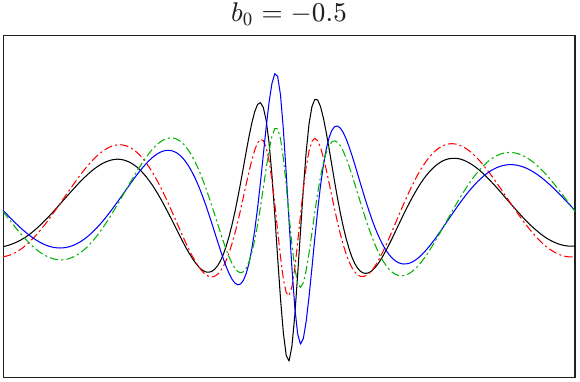}
    \par
    \vspace{0.5cm}
    \includegraphics[width=4.6cm]{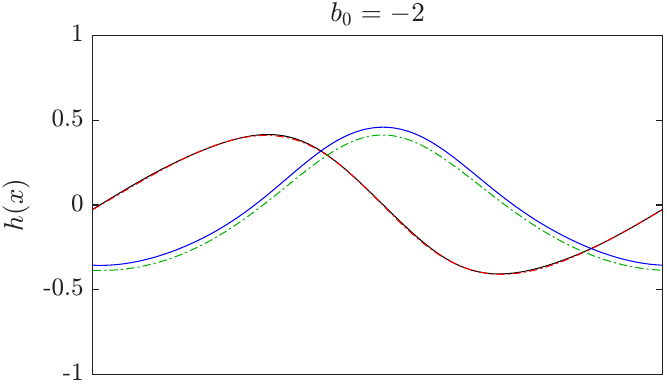}
    \hspace{0.2cm}
    \includegraphics[width=4cm]{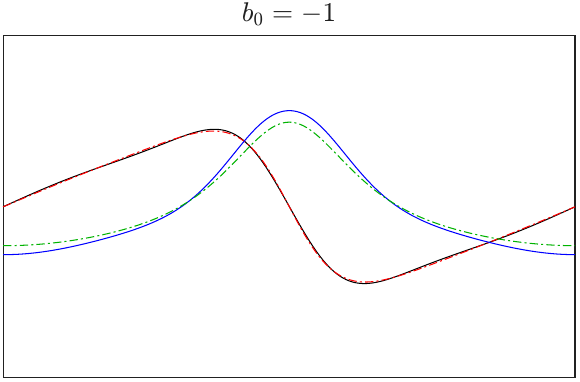}
    \hspace{0.2cm}
    \includegraphics[width=4cm]{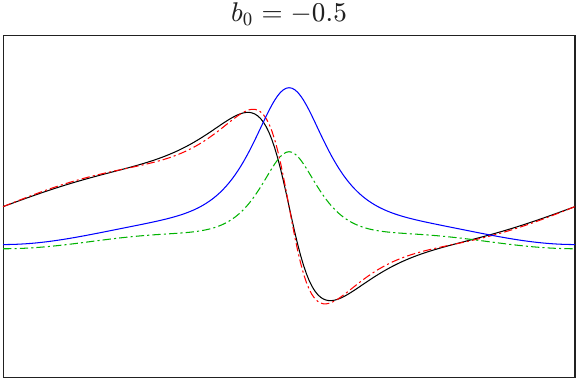}
    \par
    \vspace{0.5cm}
    \includegraphics[width=4.7cm, height=3.2cm]{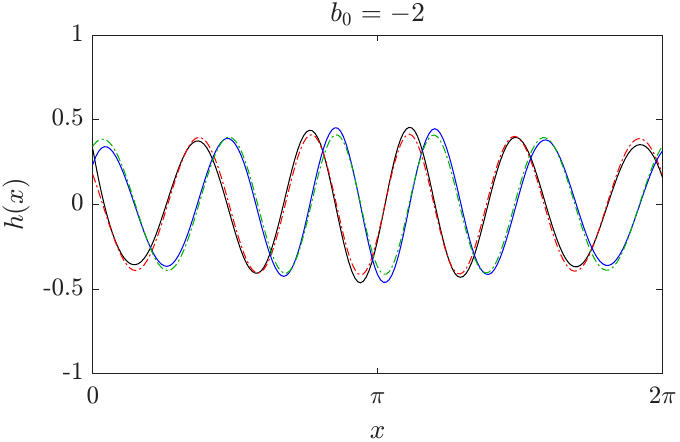}
    \hspace{0.15cm}
    \includegraphics[width=4.05cm, height=3.2cm]{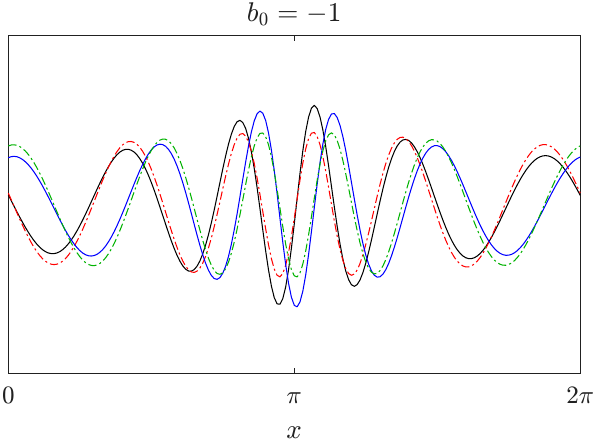}
    \hspace{0.15cm}
    \includegraphics[width=4.05cm, height=3.2cm]{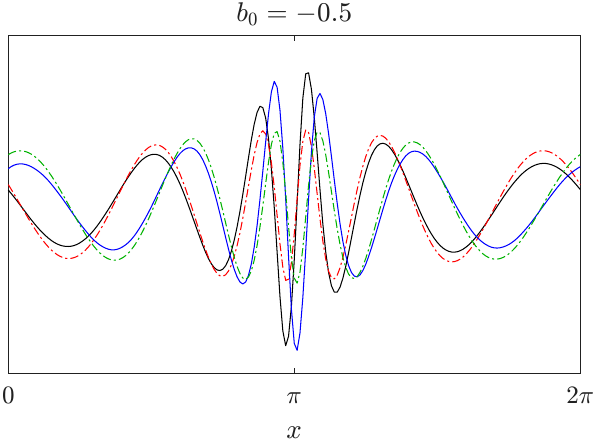}
    \caption{Comparison of full-depth Euler and surface-only model eigenfunctions for $\BS = -2, -1, -0.5$. Plotted are the real and imaginary parts of the free surface elevation $h(x)$. 
    Real part Euler (black full line); real part model (red-dashed line); imaginary part Euler (blue full line); imaginary part model (green-dotted line).
    First row: mode $0$; second row: mode $1$; third row: mode $2$; fourth row: mode $3$; fifth row: mode $4$; sixth row: mode $5$.  An animation of the results for $\BS = -0.5$ is given in Movie~1 in the supplementary material.
    }
    \label{fig:comparison_eigenfunctions_numerics_Vs_model}
\end{figure}%

To further validate the accuracy of our reduced surface-only model, we plot in figure~\ref{fig:comparison_eigenfunctions_numerics_Vs_model} the perturbation height $h(x)$ of the eigenmodes corresponding to the eigenvalues $\omega_0, \dots, \omega_5$. The phase and amplitude of each eigenmode have been chosen to make the real and imaginary parts of the eigenfunction as independent as possible, and to give each eigenmode an $L_2$ norm of $1$. As with the eigenvalues, the agreement between the two sets of eigenmodes is remarkably good. Even for the substantially deformed free surface at $\BS = -0.5$, where visible differences appear between the two sets of eigenfunctions, the surface-only eigenfunctions still capture the features of the full-depth eigenfunctions very well.

The eigenmode corresponding to $\omega_0=0$ is a Goldstone mode \citep{goldstone1961field}, arising from the broken translational symmetry of the base flow. While the governing equations are translationally invariant, the steady base flow breaks this continuous symmetry. An infinitesimal translation of the base flow generates another physically equivalent steady base state, and hence the corresponding infinitesimal perturbation is neutrally stable. In our case, the Goldstone mode has eigenvalue $\omega_0=0$, with eigenmode given by $(\upert, \wpert, p, h) = \partial_x(U_{0}, W_{0}, P_{0}, H_{0})$

Due to large deformations of the base state as $\BS$ approaches zero, we are able to compute eigenmodes from the full-depth stability equations reliably only up to $\BS = -0.3$. However, we can then use our reduced surface-only model to predict the trend of the eigenvalues as $\BS$ gets closer to zero, where a cusp-shaped base free surface forms (the maximum value of $\BS$ considered here being $\BS = -0.1$). The continuation of the eigenvalues branches within this extreme parameter limit is shown in panel~(b) of figure~\ref{fig:comparison_evals_num_vs_model}.

\subsection{Onset of instability}
\label{sec:onset_instability}

\begin{figure}%
\centering%
\raisebox{-0.9\height}{
\includegraphics[width=0.6\textwidth]{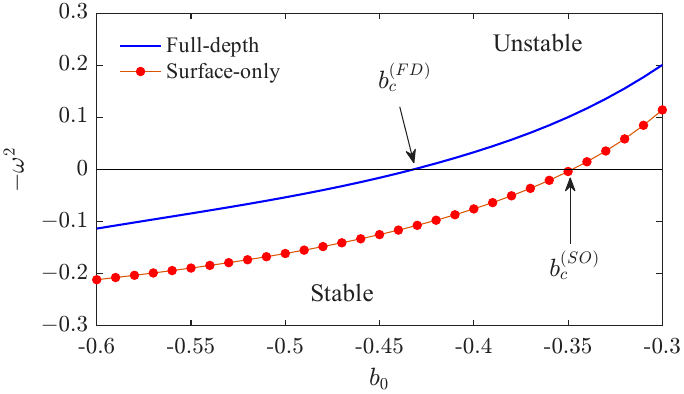}
}\par%
\caption{Instability diagram as function of $b_0$. The critical values of $b_0$ are $b^{(FD)}_{c} = -0.43$ for the 
full-depth equations and $b^{(SO)}_{c} = -0.35$ for the surface-only model. For $b_0 < b_{c}$, the system is stable; for $b_0 > b_{c}$, the system is unstable.
}%
\label{fig:onset_instability_eigenvalues}%
\end{figure}

The lowest-frequency modes $\omega_{\pm 1}$ become unstable at the critical values $b_0 = b^{(FD)}_{c} = -0.43$ for the full-depth equations and $b_0 = b^{(SO)}_{c} = -0.35$ for the surface-only model equations. In figure~\ref{fig:onset_instability_eigenvalues}, we show the onset of the instability by plotting $-\omega_{\pm 1}^2$ as a function of $b_0$. When $-\omega_{\pm 1}^2 < 0$, the eigenvalues $\omega_{\pm 1}$ form a real pair, corresponding to stable oscillations. When $-\omega_{\pm 1}^2 > 0$, the eigenvalues form a purely imaginary pair. We label the eigenvalue with positive imaginary part as $\omega_1 = i\sigma_1$, where $\sigma_1 > 0$, corresponding to exponential growth of the form $e^{\sigma_1 t}$. 

This scenario is characteristic of a Hamiltonian pitchfork bifurcation
\citep{marsden1992lectures}. Let 
$$
\mu = \frac{b_0-b_c}{|b_c|}
$$
be the reduced bifurcation parameter measuring the relative distance from the critical point. Then, in the vicinity of the critical point $\mu=0$, the eigenvalues behave as 
\begin{equation}
    \omega \sim \pm\sqrt{-\mu}. 
    \label{equ:normal_form}
\end{equation}
Thus, when $b_0 < b_c$, we have $\mu < 0$, so the argument of the square root is positive and the eigenvalues are purely real. When $b_0 > b_c$, we have $\mu > 0$, so the argument of the square root is negative and the eigenvalues are purely imaginary.

There are two caveats concerning the interpretation of this bifurcation. First, we consider here only linear stability and do not investigate the nonlinear states that may emerge following the instability. We therefore do not claim to establish a Hamiltonian pitchfork bifurcation in the full nonlinear sense. Rather, we observe that the eigenvalues exhibit the standard collision scenario associated with a Hamiltonian pitchfork bifurcation. Second, at the critical point the persistent zero eigenvalue associated with the Goldstone mode coincides with the pair of eigenvalues $\omega_{\pm1}$ that become unstable. The eigenmode corresponding to the stable wave increasingly resembles the Goldstone mode as the critical point is approached: its frequency tends to zero, and at the critical point the real part of the eigenfunction coincides with the Goldstone mode. Thus, at the critical point, the zero eigenvalue has algebraic multiplicity at least three but geometric multiplicity one. The linearization at the critical point therefore has a $3\times3$ Jordan block associated with the zero eigenvalue. 

\begin{figure}
    \centering
    \includegraphics[width=0.6\textwidth]{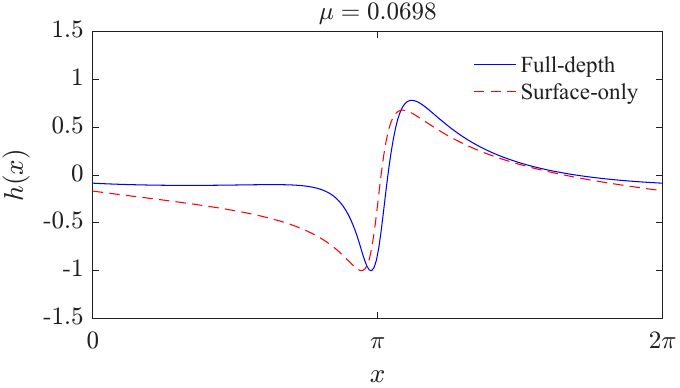}
    \caption{Comparison of the unstable eigenfunction between full-depth and surface-only model solutions computed at a bifurcation parameter $\mu = 0.0698$. This corresponds to $b_0 = -0.4$ for full-depth, and $b_0 = -0.32$ for surface-model.}
    \label{fig:onset_instability_comparison_eigenfunctions}
\end{figure}

In figure~\ref{fig:onset_instability_comparison_eigenfunctions}, we show the unstable eigenfunctions for both the full-depth and surface-only equations at the same reduced bifurcation parameter $\mu = 0.0698$, slightly above the onset of the instability. The two eigenmodes differ, but they have similar characteristics. Both resemble the Goldstone mode to some degree, although the Goldstone mode is antisymmetric about $x=\pi$, whereas neither of the unstable modes is.
The unstable mode illustrates a small issue with the surface-only model. Let $\bar h$ be the mean of $h$ over the horizontal domain. There is no constraint in the surface-only equations requiring $\bar h=0$, as is necessary for conservation of mass. For all the stable solutions considered up to this point, $\bar h$ is found to be zero. However, it is evident in figure~\ref{fig:onset_instability_comparison_eigenfunctions} that $\bar h < 0$ for the plotted surface-only eigenmode.

Figures~\ref{fig:comparison_evals_num_vs_model}--%
\ref{fig:onset_instability_comparison_eigenfunctions} demonstrate that the surface-only model reproduces very well the behaviour of the full-depth calculations as $b_0$ is varied, both in terms of the eigenvalues and the eigenfunctions. Significantly, it successfully captures the onset of the instability. This provides strong evidence that the surface-only model is capable of reproducing dynamics of linear surface waves on a non-trivial base flow with substantial vorticity and a significantly deformed free surface.

\section{Conclusions}\label{sec:conclusions}

We have generalised the spatially two-dimensional set of equations previously derived in \citet{Zuccoli2025} to study the propagation of deep-water surface waves over a general steady base free surface flow. The model 
reduces the full-depth 3D problem to a 2D problem in horizontal coordinates only,  
which enables the computation of surface waves propagating on general depth-dependent free surface flows (including non-potential flows) whose steady base free surface is not flat.
The surface-model is obtained by evaluating the linearized Euler equations on the base free surface and then introducing a closure equation to account for the vertical derivatives at the free surface.
These surface-only equations have been derived
from a heuristic argument following the same idea introduced in \citet{Zuccoli2025}.

We have verified the validity and the accuracy of the reduced surface-model to capture deep-water surface modes for two test problems of increasing complexities. 
The first test has been performed against a uni-directional, depth-dependent base flow confined in a layer of constant depth $\hinfty$. The flow is characterized by a reference velocity $\mathcal{U}$ at the undisturbed free surface $z = 0$, and by a reference horizontal length $\lambda$.
Although the free surface deformation in this case is null, the resulting exact stability problem cannot be solved by separation of variables, thus leading to a non-trivial structure in the corresponding eigensolutions.
We have shown that the model is capable of correctly reproduce both the eigenvalues and the eigenfunctions of the exact stability calculations up to moderate dimensionless flow speeds given by the Froude number $\mathcal{U}/\sqrt{g\lambda} = F \in \{0.25, 0.5, 1\}$. Moreover, we have shown the range of validity of our reduced model as function of the water depth, and shown that good agreement is reached even down to an intermediate-deep water limit $\hinfty/\lambda \gtrsim 1$.

The second test problem has been conducted against a well-known analytical solution of the fully-nonlinear free surface steady Euler equations, namely the Gerstner waves solution. The flow associated to such waves is defined in Lagrangian coordinates and extends indefinitely below an upper non-flat free surface located at $z = H_{0}(x)$.
The base flow used in our computations is defined by a one parameter family of solutions dependent on $k\BS$, where $k$ is the wavenumber of the Gertsner wave and parameter $\BS \le 0$ adjusts the steepness of the base free surface elevation.
We linearized the fully-nonlinear governing equations and boundary conditions around this background steady flow and study the resulting exact stability problem varying $k\BS$.
Even in this case the agreement between the exact and the model computations is remarkable both qualitatively and quantitatively for $k\BS \lesssim -1$. From $k\BS > -1$ the quantitative predictions of the surface-model starts to deviate more significantly, although qualitatively that is still able to follow the same trend as the
full-depth calculations. For example, the accurate prediction of the collapse of the eigenvalues associated with the first mode and the onset of the corresponding instability (figures \ref{fig:comparison_evals_num_vs_model}(a) and \ref{fig:onset_instability_eigenvalues}), and the exact emergence of the Goldstone mode in the system are notable.  In fact, as $k\BS$ approaches zero nonlinearity becomes progressively more important, and the deviation between our closure model and the full linearized Euler equations will become less significant than the error caused by the lack of nonlinearity in the linearized Euler equations.

Our study finds for the first time, to the best of our knowledge, that sufficiently steep Gerstner waves are unstable to two-dimensional perturbations.
The stability of Gerstner waves was previously studied in three-dimensions under a short-wavelength assumption using a Lagrangian formalism \citep{leblanc2004local}. Future research could extend our two-dimensional investigation using the closure condition presented here to study three-dimensional perturbations to Gerstner waves without making the short-wavelength assumption.  Also, because of our use of periodic boundary conditions, our modelling here does not allow very long wavelength perturbations, longer than the Gerstner wave wavelength. To analyse this would require a Floquet--Bloch analysis, which we leave for future work.

It is worth mentioning that our fully general reduced surface-model could also be used to make predictions of the behavior of deep-water surface waves on complex flows with complex surface deformations that are out of reach with fully three-dimensional simulations, such as a vortex dipole, as briefly discussed by \citet{vivanco2004experimental} from an experimental point of view, or even oblique Gerstner-type waves.
Other extensions to the present work could include the generalization of the reduced model derived in this work to account for other physical mechanisms and limits, such as the inclusion of capillary and dissipative effects, and its extension to weakly nonlinear and fully-nonlinear water waves.
Finally, the model is a time-dependent, two-dimensional system that is not restricted to describing just eigenmode solutions, and could be used in time-domain simulations. We envision it being used to investigate phenomena that would be extremely challenging to simulate using the full time-dependent Euler equations, for example, the scattering
of surface waves by strong vortices in deep water or the time-dependent propagation of highly dispersive water waves, or even the interaction of surface waves with meta-material-like periodic arrays generated by periodic base flows.

\begin{acknowledgements}
\noindent\textbf{Supplementary material.}
Animations of the $\BS=-0.5$ results of figure~\ref{fig:comparison_eigenfunctions_numerics_Vs_model} are available in the supplementary material as Movie~1.

\vspace{1ex}
\noindent\textbf{Acknowledgements.}
For the purpose of open access, the authors have applied a Creative Commons Attribution (CC
BY) licence to any Author Accepted Manuscript version arising from this submission.

\vspace{1ex}
\noindent\textbf{Funding.}
E.J.B.\ gratefully acknowledges the support of the UK Research and Innovation Future Leaders' Fellowship (UKRI grant MR/V02261X/1).

\vspace{1ex}
\noindent\textbf{Declaration of interests.}
The authors report no conflict of interest.
\end{acknowledgements}

\renewcommand{\theHsection}{A\arabic{section}}
\appendix  
\section{Detailed derivation and analysis of the numerical solution applied to test problem 1}\label{sec:appendix_numerics}

\subsection{Numerical solution of the eigenvalue problem}

We solve equations \eqref{equ:complete_normal_modes_pb2} using a spectral collocation method. We expand the modal pressure $\phi(y, z)$ in Fourier and Chebyshev components along the spanwise and the vertical directions, respectively \citep{trefethen2000}. By using Fourier nodes along the $y-$coordinates, we ensure periodic boundary conditions are automatically satisfied at numerical level. To account for the vertical dependence and boundary conditions along the free surface using Chebyshev nodes, it is first necessary to remap the vertical domain from the physical one to $[-1, 1]$. This is done by employing the following maps, dependent on whether the perturbation problem is solved in a finite-deep water domain ($h_\infty<\infty$), or in an infinitely deep-water one ($h_\infty=\infty$):
\begin{subequations}
    \begin{align} &
        z(z_c) = \frac{h_\infty}{2}(z_c - 1), \quad h_\infty < \infty, \\ &
        z(z_c) = B_{\mathrm{map}}\tan\Big[\frac{\pi}{4}(z_c-1)\Big], \quad h_\infty = \infty,
    \end{align}
    \label{equ:remaps_test_problem1}
\end{subequations}
with $z_c$ denoting the Chebyshev discretization points in $[-1, 1]$. 
The vertical gradient transforms according to
\begin{subequations}
    \begin{align} &
        \frac{\partial}{\partial z} = \frac{2}{h_\infty}\frac{\partial}{\partial {z_c}}, \quad h_\infty < \infty, \\ &
        \frac{\partial}{\partial z} = \Big(\frac{4}{\pi B_{\mathrm{map}}}\Big) \frac{1}{1 + \tan^{2}\Big[\frac{\pi}{4}(z_c - 1)\Big]}\frac{\partial}{\partial {z_c}}, \quad h_\infty = \infty.
    \end{align}%
    \label{equ:axial_gradient_transformations}%
\end{subequations}%
Calling $D_y, D_z, D^{2}_y, D^{2}_z$ the resulting numerical differentiation matrices in the two spatial directions, and $\Phi_{i,j} = \phi(y_i, z_{j})$--$(i,j) \in [1, N_y]\times[1, N_z]$--the values of the eigenfunction $\phi$ at the spectral nodes, equations \eqref{equ:complete_normal_modes_pb2} at numerical level become
\begin{subequations}
    \begin{align} &
        \Big[(Fk\Zbase(z_{j})\Ybase(y_i) - \omega)(D^{2}_y + D^{2}_z - k^2)
        -2Fk\Zbase(z_j)\Ybase'(y_i)D_y
        -2Fk\Zbase'(z_j)\Ybase(y_i)D_z\Big]\Phi_{i,j} = 0, \label{equ:numerical_governing_eq} \\ &
        [D_z -\omega^2 + 2\omega Fk\Ybase(y_i) - F^2k^2q^{2}_0(y_i)]\Phi_{i,1} = 0, \label{equ:numerical_FSBC} \\ &
        [D_z]\Phi_{i,N_z} = 0. \label{equ:numerical_impermeable_BC}
    \end{align}%
    \label{equ:complete_normal_modes_pb2_numerical}%
\end{subequations}%
The boundary conditions \eqref{equ:numerical_impermeable_BC} and \eqref{equ:numerical_FSBC} are enforced directly in the bulk equation \eqref{equ:numerical_governing_eq} by finding the indices of the bulk matrices where $z_c = -1$ and $z_c = 1$, respectively, and substitute the corresponding discretized operators contained in the numerical boundary conditions. This yields a polynomial eigenvalue problem which can be compactly written in matrix form as
\begin{equation}
    [\vect{K_{N}} + \omega \vect{C_N} + \omega^2 \vect{M_{N}}]\vect{\Phi}_{N} = \vect{0},
    \label{equ:compact_poly_eig_problem}
\end{equation}
where $\vect{K_{N}}, \vect{C_{N}}, \vect{M_{N}}$ represent the stiffness, damping and mass matrices coming from the implementation of equations \eqref{equ:complete_normal_modes_pb2_numerical}, and $\vect{\Phi}_{N} = (\Phi_{1, 1}, \dots, \Phi_{1, N_z}, \Phi_{2, 1}, \dots, \Phi_{2, N_z}, \dots)$ is the total array of the eigenfunction $\phi(y,z)$ evaluated at the spatial nodes.

\subsection{Spurious numerical modes and resolvedness conditions}

It is well known that spectral methods are prone to giving rise to spurious eigenvalues and eigenfunctions.
These are numerical artifacts caused by numerical under-resolution of highly-oscillatory modes; that is, eigenfunctions that do satisfy the discretized problem but do not approximate solutions of the continuous eigenvalue problem. Following~\citet{brambley+peake-2008} \citep[see also][p.~58--61]{brambley-2007}, we implement two tests to remove such spurious numerical modes from the numerical solution of equation \eqref{equ:compact_poly_eig_problem}. This is a modified version of the numerical tests described in appendix~A of \citet{zuccoli+brambley+barkley-2022} to account for the presence of doubly-degenerate eigenvalues in the present case.  We reproduce these tests here in their entirity to save the reader from having to refer elsewhere for the details.

\begin{figure}
    \centering
    \includegraphics[width=10cm]{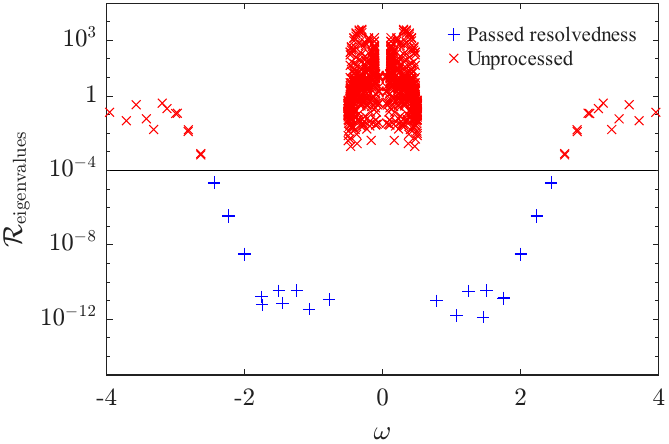}
    \caption{Graph showing the continuity of eigenvalues for $\Fr = 0.5$, $k = 1$, $N_y = 26$, $N_z = 36$, compared against $N_y = 30$, $N_z = 40$ with the same settings. The tolerance threshold $\mathrm{tol} = 10^{-4}$ is displayed with a full black line.}
    \label{fig:continuity_evals_TEST1}
\end{figure}

The first condition (continuity of the eigenvalues) requires that the eigenvalues do not move significantly when the numerical discretization is slightly changed. Suppose eigenvalues $\{\omega_{p}\}$ have been computed using a discretization $(N_y, N_z)$, and suppose $\{\hat{\omega}_{q}\}$ have been computed with a slightly increased resolution, such as $(N_{y}+4, N_{z}+4)$. The first condition is then given by requiring that, for each eigenvalue $\omega_p$,
\begin{equation}
\mathcal{R}_{\mathrm{eigenvalues}} = \frac{\inf_q|\omega_{p} - \hat{\omega}_{q}|}{\sqrt{d^{2}_1 + d^{2}_2}} \le \text{tol},
\label{resolv_cond_eigenvalues}
\end{equation}
where $d_i$ is the distance between $\omega_p$ and the $i^{\mathrm{th}}$ nearest $\omega_{\ell}$ with $\ell\neq p$, and $\text{tol}$ is a prescribed tolerance. In other words, $\mathcal{R}_{\mathrm{eigenvalues}}$ is the distance moved by an eigenvalue between the two resolutions, normalized by the distance to the next closest eigenvalue so as to be scale free. The reason for the averaging over $d_1$ and $d_2$ is to deal with doubly-degenerate eigenvalues. A small resolvedness on an eigenvalue continuity indicates the eigenvalue is stable under small changes to the discretization, and hence that it is a good candidate for a physical eigenvalue.
Figure \ref{fig:continuity_evals_TEST1} shows the resolvedness test on the eigenvalues computed for $\Fr = 0.5$, $k = 1$, and for spatial resolutions $(N_y, N_z) = (26, 36)$ on one hand, and  $(N_y, N_z) = (30, 40)$ on the other. The tolerance has been set to $10^{-4}$.

The second resolvedness condition involves the spectral coefficients of the numerical eigenvectors. In particular, we require the modulus of the spectral coefficients to decay smoothly as the number of grid points $N_y$ and $N_z$ gets larger.
So, let $\hat{\phi}_{ij}$ be a generic spectral coefficient, the following condition has been implemented
\begin{equation}
\begin{aligned} &
\mathcal{R}_{\mathrm{eigenvectors}} = \frac{\sup_{(i, j) \in \mathcal{B}} |\hat{\phi}_{ij}|}{\sup_{\mathrm{all}\,i, j} |\hat{\phi}_{ij}|} \le \text{tol}, 
\end{aligned}
\label{resolvedness}
\end{equation}
where $\mathcal{B}$ is a part of the domain in the $N_{y}-N_{z}$ plane, defined as
\begin{equation}
\mathcal{B} = \big\{ (i, j): N_{y}/2 - b_{y} \le |i| \le N_{y}/2 \quad \text{ or } \quad N_{z} - b_{z} \le j \le N_{z} \big\},
\end{equation}
$\text{tol}$ is a prescribed tolerance on the magnitude of the coefficients and $b_y, b_z$ two prescribed borders widths in the $N_{y}-N_{z}$ plane.
\begin{figure}
    \centering%
    \includegraphics[width=12cm]{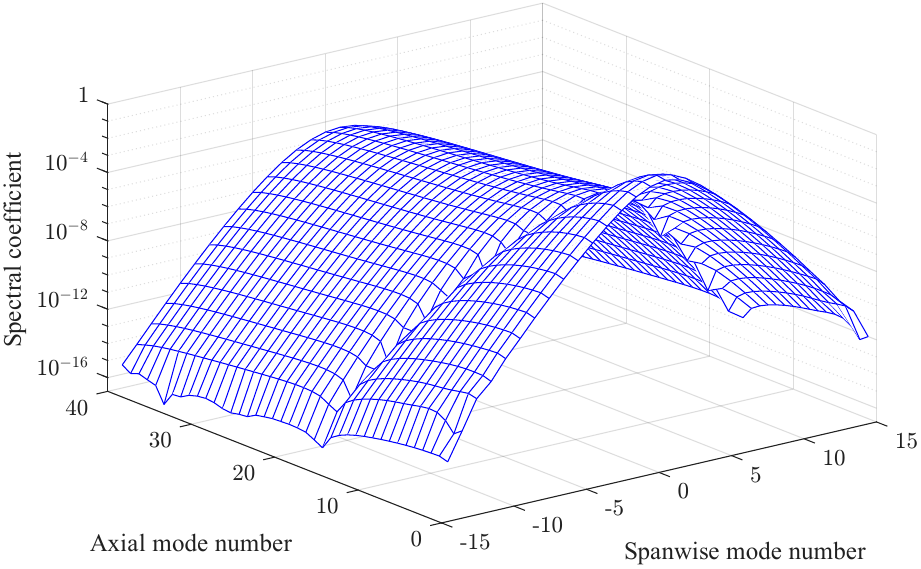}
    \par
    \vspace{0.5cm}
    \includegraphics[width=12cm]{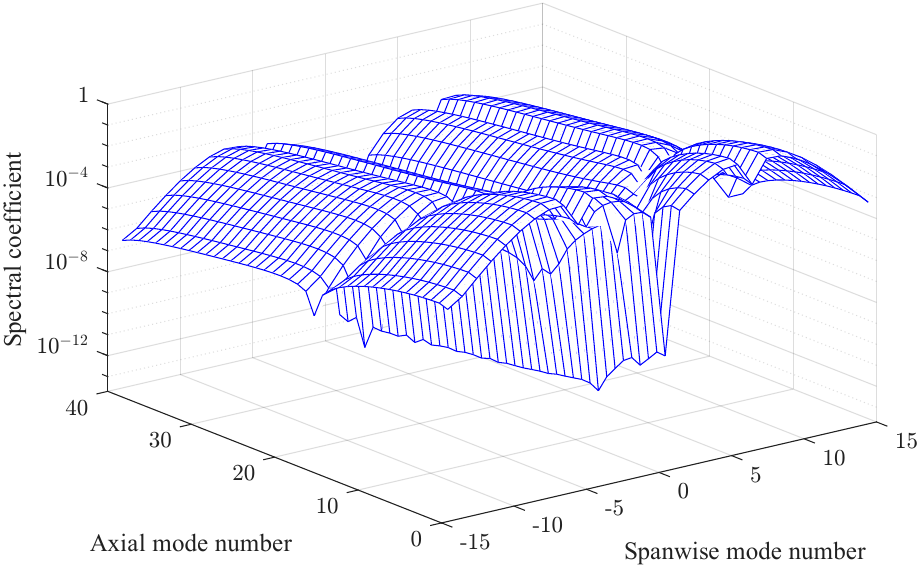}
    \caption{Example of a well-resolved mode (Top) and a poorly-resolved mode (Bottom) for $\Fr = 0.5$, $k = 1$, and spatial resolutions $(N_y, N_z) = (26, 36)$, against $(N_y, N_z) = (30, 40)$.}
    \label{fig:well_bad_resolved_modes}
\end{figure}
Figure~\ref{fig:well_bad_resolved_modes} compares the magnitude of the spectral coefficients for a well resolved mode and a poorly-resolved mode for $N_{y} = 52$, $N_{z} = 100$, $b_y = b_z = 4$, and both tolerances $\text{tol} = 10^{-4}$. The well-resolved mode can be seen to have spectral coefficient $|\phi_{ij}|$ decreasing exponentially quickly. By contrast, for the poorly-resolved mode, the spectral coefficient $|\phi_{ij}|$ is unstable as function of the spatial resolution; hence it is discarded.

\section{Exact solutions of the eigenvalue problems in the case of a flat free surface}\label{sec:app_exact_solutions_flat_FS}

In this section we want to obtain the exact eigensolutions for both the first model validation problem and the Gerstner wave perturbation problem in the limit the steady free surface becomes flat. In the first case this condition requires $\Fr = 0$; in the second it requires $\BS \rightarrow \infty$. For the purposes of this paper, we show here the solutions in the infinitely deep-water regime.
\subsection{Exact solution validation test 1}
\label{sec:appendix_model1}

When $\Fr = 0$ and $h_\infty = \infty$, the eigenproblem \eqref{equ:complete_normal_modes_pb2} simplifies into
\begin{subequations}
    \begin{align} &
        \frac{\partial^2\phi}{\partial y^2} + \frac{\partial^2\phi}{\partial z^2} - k^2 \phi = 0, \\ &
        \frac{\partial\phi}{\partial z} = \omega^2\phi, \quad \mathrm{on} \quad z = 0, \\ &
        \phi\rightarrow 0, \quad \mathrm{as} \quad z \rightarrow -\infty.
    \end{align}%
    \label{equ:eigenPB_TEST1_flatFS}%
\end{subequations}%
Due to the periodicity in $y$, we can search for a solution in the form $\phi(y, z) = f(z)e^{\I n y}$, with $n \in \mathbb{Z}$. This gives the eigenproblem
\begin{subequations}
    \begin{align} &
        f'' - (n^2 + k^2) f = 0, \\ &
        f'(0) = \omega^2 f(0), \quad \mathrm{and} \quad f(-\infty) = 0,
    \end{align}%
    \label{equ:eigenPB_TEST1_flatFS_reduced}
\end{subequations}%
whose solutions read
\begin{equation}
    f(z) = e^{\sqrt{n^2 + k^2}z}, \quad \mathrm{and} \quad
    \omega_{n} = \pm\sqrt[4]{n^2 + k^2}, \quad n \in \mathbb{Z}.
    \label{equ:exact_solutions_TEST1_flatFS}
\end{equation}

\subsection{Exact solution validation test 2}
\label{sec:appendix_gerstner}

When $\BS = -\infty$, the velocity is uniform in the $x$-direction only and all the quantities depending on $\BS$ goes to zero. Hence, the eigenproblem \eqref{equ:generalized_eval_PB_Gerstner} can be re-written entirely in terms of the pressure and assumes the following form
\begin{subequations}
    \begin{align} &
        \frac{\partial^2\hat{p}}{\partial x^2} + \frac{\partial^2\hat{p}}{\partial z^2} = 0, \\ &
        \frac{\partial\hat{p}}{\partial z} = -\Big(\frac{\partial}{\partial x}-\I\omega \Big)^2\hat{p}, \quad \mathrm{on} \quad z = 0, \\ &
        \hat{p}\rightarrow 0, \quad \mathrm{as} \quad z \rightarrow -\infty.
    \end{align}%
    \label{equ:eigenPB_TEST2_flatFS}%
\end{subequations}%
Again, due to the periodicity in $x$, we can search for a solution in the form $\hat{p}(y, z) = \psi(z)e^{\I n x}$, with $n$ being an integer. This returns the eigenproblem
\begin{subequations}
    \begin{align} &
        \psi'' - n^2 \psi = 0, \\ &
        \psi'(0) = (\omega - n)^2 \psi(0), \quad \mathrm{and} \quad \psi(-\infty) = 0,
    \end{align}%
    \label{equ:eigenPB_TEST2_flatFS_reduced}%
\end{subequations}%
whose solutions read
\begin{equation}
    \psi(z) = e^{|n|z}, \quad \mathrm{and} \quad
    \omega_{n} = n\pm\sqrt{|n|}, \quad n \in \mathbb{Z}\setminus\{0\}.
    \label{equ:exact_solutions_TEST2_flatFS}
\end{equation}
Note that the convention used to denote the eigenvalues computed numerically and shown in figure \ref{fig:comparison_evals_num_vs_model} is that of increasing modulus, i.e. $|\omega_0| < |\omega_{\pm 1}| < |\omega_{\pm 2}| \dots$ and does not correspond to the notation used in equation \eqref{equ:exact_solutions_TEST2_flatFS}, which instead denotes the eigenvalues based on the periodicity of the corresponding eigenfunctions.

\section{Unphysical modes of the model}
\label{sec:unphysical_modes_model}

The general form of the surface-only model equations employed and tested in this work is given by equations \eqref{equ:reduced_system_Gerst}. 
Seeking normal modes of the form $e^{-\I\omega t}$, equations \eqref{equ:reduced_system_Gerst} give rise to a generalised eigenvalue problem of the form
\begin{subequations}\begin{align} &
U_{0}|_{H_0}\partial_x u + 
u \left. \partial_x U_0\right|_{H_{0}} + 
w \left. \partial_z U_0\right|_{H_{0}}
+ \partial_x(g_0 h) - 
H_0' \left.{\frac{\partial p}{\partial z}}\right|_{H_{0}} = \I\omega u, \\ &
U_{0}|_{H_0}\partial_x w +
u \left. \partial_x W_0\right|_{H_{0}} + 
w \left. \partial_z W_0\right|_{H_{0}}
+ \left.{\frac{\partial p}{\partial z}}\right|_{H_{0}}  = \I\omega w,  \\ &
U_{0}|_{H_0}\partial_x\Big(\left.{\frac{1}{g_0}\frac{\partial p}{\partial z}}\right|_{H_{0}}\Big) - \frac{\varphi_0}{g_0}\left.{\frac{\partial p}{\partial z}}\right|_{H_{0}} + 
\partial_x u
= \I\omega \left.{\frac{1}{g_0}\frac{\partial p}{\partial z}}\right|_{H_{0}}, \\ &
U_{0}|_{H_0}\partial_x h
+ u H_0'
- \varphi_{0}h - w = \I\omega h, 
\end{align}%
\label{equ:reduced_system_Gerst_evalPb}%
\end{subequations}
with $\omega$ being the eigenvalues and $\vect{\hat{w}} = (u, w, \partial_z p|_{H_0}, h)^{T}$ the array of the eigenfunctions. 
For validation test 1, equations \eqref{equ:reduced_system_Gerst_evalPb} simply reduce to
\begin{subequations}\begin{align} &
U_{0}|_{H_0}\partial_x u + 
w \left. \partial_z U_0\right|_{H_{0}}
+ \partial_x h = \I\omega u, \\ &
U_{0}|_{H_0}\partial_x w + \left.{\frac{\partial p}{\partial z}}\right|_{H_{0}}  = \I\omega w,  \\ &
U_{0}|_{H_0}\partial_x\Big(\left.{\frac{\partial p}{\partial z}}\right|_{H_{0}}\Big) + 
\partial_x u
= \I\omega \left.{\frac{\partial p}{\partial z}}\right|_{H_{0}}, \\ &
U_{0}|_{H_0}\partial_x h - w = \I\omega h, 
\end{align}%
\label{equ:reduced_system_model1}%
\end{subequations}
where now $H_0 = 0$ and $U_0 = \Fr f_0(z)q_0(y)$, as defined in \eqref{eq:base_test1}.

It is found that the full spectrum of the surface-model equations contains some unphysical eigenvalues that need to be removed. Such unphysical modes arise because the surface-model equations cannot actually distinguish whether the fluid extends infinitely deep below or infinitely deep above the base free surface $H_0(x)$.

In order to decide which eigenvalue is good and which needs to be discarded, we define the following parameter 
\begin{equation}
    \lambda = \int_{0}^{2\pi}\mathrm{Re}\Big(\frac{\partial_{z}p|_{H_0}}{p|_{H_0}}\Big)\mathrm{d}x = \int_{0}^{2\pi}\mathrm{Re}\Big(\frac{\partial_{z}p|_{H_0}}{g_0 h}\Big)\mathrm{d}x,
\end{equation}
where the integration variable $x$ here should be interpreter as either $y$ for the validation test 1, and as $x$ for the validation test 2.
Parameter $\lambda$ can be interpreted as a mean value of the axial decay/growing rate of the modes. Thus, being us interested in the propagation of surface waves above an infinitely extended fluid, if $\lambda$ is sufficiently positive, the mode is likely to be good and physical. Viceversa, if $\lambda$ is sufficiently negative, the mode is likely to be unphysical and needs to be discarded. More quantitatively, we set a tolerance $\delta_{\lambda}$ and define the following criterion to recognise physical meaningful eigenvalues from unphysical ones:
\begin{equation}
\begin{aligned} &
    \mathrm{if}\quad \lambda > \delta_{\lambda} && \Rightarrow && \mathrm{physical\hspace{0.2cm} eigenvalue}, \\ &
    \mathrm{if}\quad  \lambda < -\delta_{\lambda} && \Rightarrow && \mathrm{unphysical\hspace{0.2cm} eigenvalue}
\end{aligned}
\label{equ:selection_criterion_evals_models}
\end{equation}
For test validation 1, the expression of $\lambda$ is obtained analytically and reads
\begin{equation}
    \lambda = \pi(2\omega^2 + k^2\Fr^2).
\end{equation}

\begin{figure}
    \centering
    \includegraphics[width=0.6\linewidth]{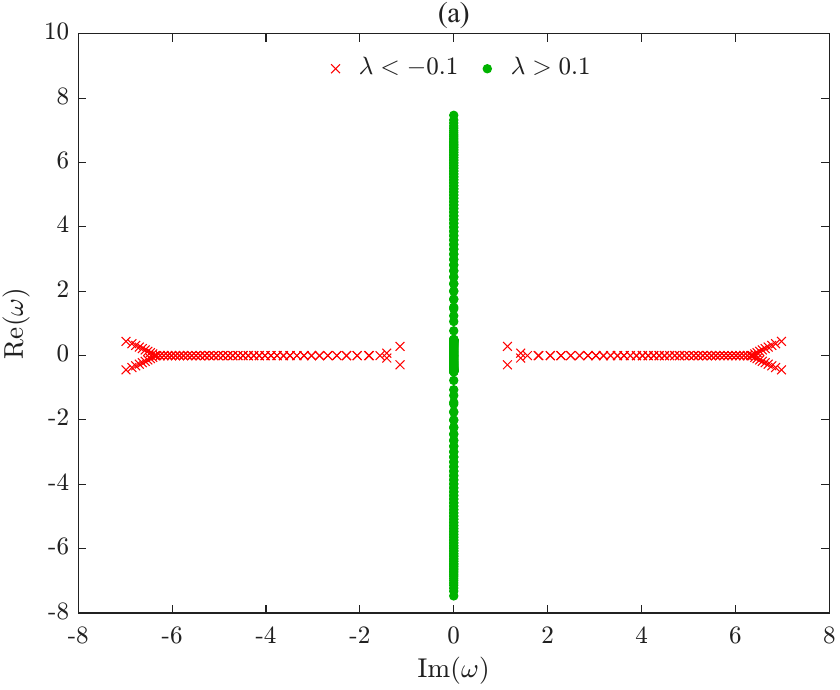}
    \par
    \vspace{0.7cm}
    \includegraphics[width=0.6\linewidth]{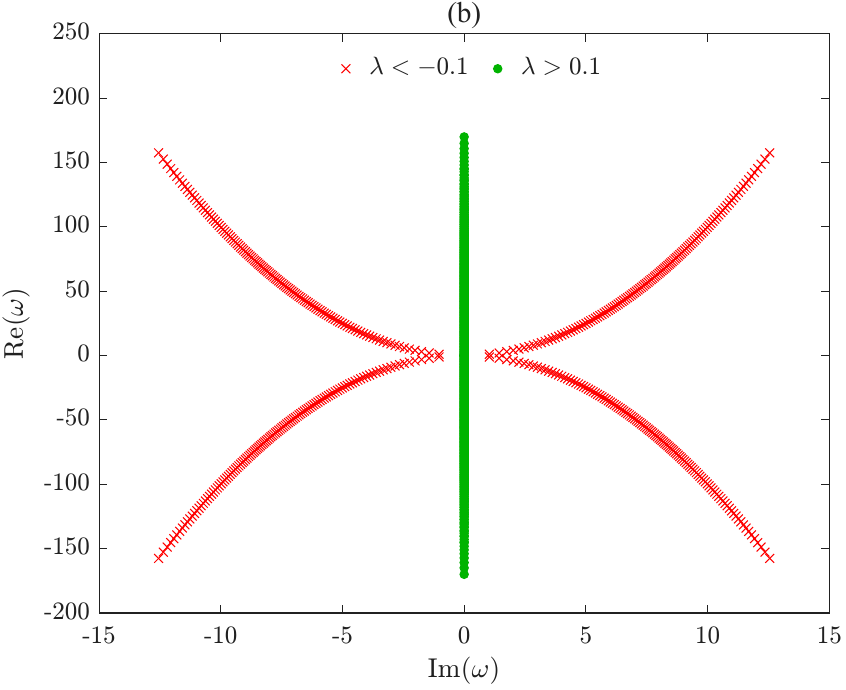}
    \caption{(a): Full spectrum of the model equations \eqref{equ:reduced_system_model1} at $\Fr = 0.5$. (b): Full spectrum of the model equations \eqref{equ:reduced_system_Gerst_evalPb} at $b_0 = -1$. Red crosses: unphysical eigenvalues not passing the selection criterion~\eqref{equ:selection_criterion_evals_models}. Green dots: physical eigenvalues passing the selection criterion~\eqref{equ:selection_criterion_evals_models}.}
    \label{fig:unphysical_evals_models}
\end{figure}

In figure \eqref{fig:unphysical_evals_models} we show the whole spectrum of the surface-only model equations \eqref{equ:reduced_system_model1}--\eqref{equ:reduced_system_Gerst_evalPb} computed at $\Fr = 0.5$ for the validation test 1, and at $b_0 = -1$ for the validation test 2, and with tolerance $\delta_{\lambda} = 0.1$, comprising  those eigenvalues passing the selection criterion (green dots) and those not passing it (red crosses). As expected for the flow parameters used here, no unstable modes are present and the physically acceptable eigenvalues are purely real. 

The parabolic branches in panel~(b) of figure~\ref{fig:unphysical_evals_models} can be thought, in fact, to be the continuation as $b_0$ gets larger of the unphysical branches obtained when $b_0 \rightarrow -\infty$. In this limit, the eigenvalues predicted by the model are given by
\begin{equation}
    \omega^{(1,2)}_{n} = n \pm \sqrt{|n|}, \quad \mathrm{and} \quad 
    \omega^{(3,4)}_{n} = n \pm \I\sqrt{|n|}, \quad n \in \mathbb{Z}\setminus\{0\}.
\end{equation}
The two branches $\omega^{(3, 4)}$ correspond to unphysical eigenvalues for which $\lambda = -1$.

\bibliography{biblio}

\end{document}